\documentclass[letterpaper,english,aps,pre,twocolumn,showkeys,nofootinbib,superscriptaddress,showpacs,floatfix]{revtex4-1}
\usepackage{color}
\usepackage{mathpazo}
\usepackage[T1]{fontenc}
\usepackage[latin9]{inputenc}
\usepackage{textcomp}
\usepackage{amsmath}
\usepackage{amssymb}
\usepackage{graphicx}
\usepackage{esint}

\makeatletter

\@ifundefined{textcolor}{}
{%
 \definecolor{BLACK}{gray}{0}
 \definecolor{WHITE}{gray}{1}
 \definecolor{RED}{rgb}{1,0,0}
 \definecolor{GREEN}{rgb}{0,1,0}
 \definecolor{BLUE}{rgb}{0,0,1}
 \definecolor{CYAN}{cmyk}{1,0,0,0}
 \definecolor{MAGENTA}{cmyk}{0,1,0,0}
 \definecolor{YELLOW}{cmyk}{0,0,1,0}
 }

\newcommand{\iR}{\mathbb{R}}
\newcommand{\iO}{\Omega}

\newcommand{\refFigEx}[1]{Fig.~\ref{fig:Rate-Estimate}\expandafter\@slowromancap\romannumeral#1@}
\newcommand{\refFig}[1]{Fig.~\ref{#1}}

\makeatother

\usepackage{babel}
\begin{document}

\title{Spectral rate theory for projected two-state kinetics}

\author{Jan-Hendrik Prinz}

\affiliation{DFG Research Center Matheon, Free University Berlin, Arnimallee 6,
14195 Berlin, Germany}

\email{jan-hendrik.prinz@fu-berlin.de}

\author{John D. Chodera}

\affiliation{Computational Biology Program, Memorial Sloan-Kettering Cancer Center,
New York, NY 10065, USA}

\email{choderaj@mskcc.org}

\author{Frank No\'e}

\email{frank.noe@fu-berlin.de (corresponding author)}

\affiliation{DFG Research Center Matheon, Free University Berlin, Arnimallee 6,
14195 Berlin, Germany}
\begin{abstract}
Classical rate theories often fail in cases where the observable(s) or order
parameter(s) used are poor reaction coordinates or the observed signal
is deteriorated by noise, such that no clear separation between reactants
and products is possible. Here, we present a general spectral two-state rate theory for
ergodic dynamical systems in thermal equilibrium that explicitly takes
into account how the system is observed. The theory allows the systematic
estimation errors made by standard rate theories to be understood
and quantified. We also elucidate the connection of spectral rate
theory with the popular Markov state modeling (MSM) approach for molecular
simulation studies. An optimal rate estimator is formulated that gives
robust and unbiased results even for poor reaction coordinates and
can be applied to both computer simulations and single-molecule experiments.
No definition of a dividing surface is required. 
Another result of the theory is a 
model-free definition of the reaction coordinate quality (RCQ). The RCQ can
be bounded from below by the directly computable observation quality (OQ), thus providing a measure
allowing the RCQ to be optimized by tuning the experimental setup.
Additionally, the respective partial probability distributions can be obtained for
the reactant and product states along the observed order parameter,
even when these strongly overlap. 
The effects of both filtering (averaging) and uncorrelated noise are also examined.
The approach is demonstrated on
numerical examples and experimental single-molecule force probe data
of the p5ab RNA hairpin and the apo-myoglobin protein at low pH , here
focusing on the case of two-state kinetics.
\end{abstract}

\keywords{rate theory; dividing surface; fitting correlation function; phenomenological rate; microscopic rate}

\maketitle

The description of complex molecular motion through simple kinetic
rate theories has been a central concern of statistical physics. A
common approach, first-order rate theory, treats the relaxation kinetics
among distinct regions of configuration space by single-exponential
relaxation. There has been recent interest in estimating such rates
from trajectories of single molecules, resulting from
the recent maturation of measurement techniques able to collect
extensive traces of single molecule extensions or fluorescence events~\cite{GreenleafBlock_AnnuRev07,LapidusEatonHofrichter_PNAS00_FCS}.
When the available observable is a good reaction coordinate in that
it allows the slowly-converting states to be clearly separated (see
\refFigEx{1}, left), classical rate theories
apply and the robust estimation of transition rates is straightforward
using a variety of means~\cite{Zhou_QuartRevBiophys10_RateReview}.
However, in the case in which the slowly-converting states overlap
in the observed signal (see \refFigEx{3}, left), either due to the fact that the molecular order parameter used
is a poorly separating them, or due to large noise of the measurement
(see discussion in \cite{DudkoGrahamBest_PRL2011_FoldingBarrier,MorrisonThirumalai_PRL11_SingleMoleculePulling}),
a satisfactory theoretical description is missing and many estimators
break down. 

Most two-state rate theories and estimators are based on dividing the observed
coordinate into a reactant and a product substate and then in some
way counting transition events that cross the dividing surface. Transition
state theory measures the instantaneous flux across this surface,
which is known to overestimate the rate due to the counting of unproductive
recrossings over the dividing surface on short timescales \cite{Eyring_JCP35_TST}.

Reactive flux theory \cite{Chandler_JCP78_ReactiveFlux} proposed
to cope with this by counting a transition event only if it has succeeded
to stay on the product side after a sufficiently long lag time $\tau$.
Reactive flux theory involves derivatives of autocorrelation functions
that are numerically unreliable to evaluate \cite{ChoderaEtAl_JCP11_RateTheory}.
In practice, one therefore typically estimates the relaxation rate
\emph{via} integration or performing an exponential fit to the tail
of a suitable correlation function, such as the number correlation
function of reactants or the autocorrelation function of the experimentally
measured signal \cite{ElsonMagdeI_Biopolymers74,SkinnerWolynes_JCP78_RelaxationProcesses,Zhou_QuartRevBiophys10_RateReview}.
In order to split this relaxation rate into a forward and backward
rate constant, a clear definition of the reactant and product substates
is needed, which is difficult to achieve when these substates overlap
in the observed signal.

Markov state models (MSMs) have recently become a popular approach to producing
a simplified statistical model of complex molecular dynamics from
molecular simulations.

While applicable only when the discretization of state space succeeds in separating 
the metastable conformations, these models can be regarded as steps towards a
multistate rate theory. MSMs use a transition matrix describing the
probability a system initially found in a substate $i$ is found in
substate $j$ a lag time $\tau$ later. When the state division allows
the metastable states of the system to be distinguished \cite{ChoderaSwopePiteraDill_MMS,ChoderaEtAl_JCP07,NoeHorenkeSchutteSmith_JCP07_Metastability,PrinzEtAl_JCP10_MSM1},
the transition matrix with a sufficiently large choice of $\tau$
can be used to derive a phenomenological transition rate matrix that
accurately describes the interstate dynamics \cite{SwopePiteraSuits_JPCB108_6571}.
This is explicitly done for the two-state case in \cite{ChoderaEtAl_JCP11_RateTheory}.
It has been shown in \cite{SarichNoeSchuette_MMS09_MSMerror,PrinzEtAl_JCP10_MSM1}
that by increasing the number of substates used to partition state
space, and hence using multiple dividing surfaces instead of a single
one, these rate estimates become more precise. In the limit of infinitely
many discretization substates, the eigenfunctions of the dynamical propagator in
full phase space are exactly recovered, and the rate estimates become
exact even for $\tau\rightarrow0^{+}$ \cite{KubeWeber_JCP07_CoarseGraining}.
In practice, however, a finite choice of $\tau$ is necessary in order
to have a small systematic estimation error, especially if ``uninteresting''
degrees of freedom such as momenta or solvent coordinates are discarded.
An alternative way of estimating transition rates is by using a state
definition that is incomplete and treats the transition region implicitly
\emph{via} committor functions that may better approximate the eigenfunctions
of the dynamical propagator in this region \cite{BucheteHummer_JPCB08,SchuetteEtAl_JCP11_Milestoning,Buchner_BBA11_ProteinFoldingKinetics}.

The quality of the rate estimates in all of the above approaches relies
on the ability to separate the slowly-converting states in terms of
some dividing surface or state definition. These approaches  often break down in practice when the available observables do not permit such
a separation, i.e., when kinetically distinct states overlap in the
histogram of the observed quantity. However, such a scenario may often
arise in single-molecule experiments where the available order parameter
depends on what is experimentally observable and may not necessarily
be a good indicator of the slow kinetics. Moreover, consequences of
the measurement process may increase the overlap between states, for
example by bead diffusion in optical tweezer experiments or by shot
noise in single-molecule fluorescence measurements. In favorable situations,
the signal quality can be improved by binning the data to a coarser
timescale (often simply referred to as ``\emph{filtering}''), thus reducing the fluctuations from fast processes and
shot noise. However, the usefulness of such filtering is limited because
the time window used needs to be much shorter than the timescales
of interest --- otherwise the kinetics will be distorted. In general,
one has to deal with a situation where overlap between the slowly-converting
states is present, both theoretically and practically.

Hidden Markov Models (HMMs) \cite{Rabiner_IEEE89_HMM,McKinney_BiophysJ06_HMM-FRET,ChoderaEtAl_BiophysJ11_BHMM}
and related likelihood methods \cite{GopichSzabe_HMMFRET_JCP09} are
able to estimate transition rates even in such situations, and have
been recently successful in distinguishing overlapping states in molecules
with complex kinetics \cite{StiglerRief_Science11_CalmodulinFoldingNetwork,PirchiHaran_NatureComms11_SingleMoleculeFRET}.
However, HMMs need a probability model of the measurement process
to be defined, which can lead to biased estimates when this model
is not adequate for the data analyzed. A recent approach, the signal
pair-correlation analysis (PCA) \cite{HoffmannWoodside_Angewandte11_PairCorrelationAnalysis}
provides rate estimates without an explicit probability model, and
instead requires the definition of indicator functions on which the measured
signal can uniquely be assigned to one of the kinetically separated
states. While this is often easier to achieve than finding an appropriate
dividing surface, there is a trade-off between only using only data
that is clearly resolved to be in one state or the other (thus minimizing
the estimation bias), while avoiding discarding too much data (thus
minimizing the statistical error). Despite these slight limitations,
both HMMs and PCA are practically very useful to identify and quantify
hidden kinetics in the data. Yet, both are algorithmic approaches
rather than a rate theory.

The recent success of single-molecule experiments and the  desire 
for a robust rate estimation procedure that yields viable rate estimates even 
when highly overlapping states indicate clearly that the observed signal is a 
poor reaction coordinate highlights the need for a general and robust two-state 
rate theory for observed dynamics. 
 Here, we make an attempt towards such a general rate theory for 
stochastic dynamics that are observed on a possibly poor reaction coordinate --- 
 often  because the probed molecular order parameter is 
a poor choice, or because the measurement device creates overlap by noise-broadening 
the signal.

Our approach requires only mild assumptions to hold for the dynamics of the observed system:
First, the dynamical law governing the time-evolution of the system \emph{in its full phase space} --- 
including all positions and velocities of the entire measured construct and the surround solvent --- 
is assumed to be a time-stationary Markov process.
We also require that the system obeys microscopic detailed balance in the full phase space, 
and supports a unique stationary distribution. These mild criteria are easily satisfied by a 
great number physical systems of interest in biophysics and chemistry.

When projected onto some measured observable, the dynamics of the system 
are no longer Markovian. In addition, the observed dynamics may be subjected 
to measurement noise. As a result, the resulting signal may not be easily 
separable into kinetically distinct states by a simple dividing surface, 
something that is often required for existing rate estimation procedures 
to work well.
 
Our framework allows us to (i) evaluate the quality of existing estimators 
and propose optimal estimators for the slowest relaxation rate, (ii) provide 
a model-free definition of the reaction coordinate quality (RCQ) and the observation quality (OQ) of the signal, 
and (iii) derive an optimal estimator for the transition rates between the slowly
converting states, as well as their stationary probability densities,
even if these strongly overlap in the observation.

The present rate theory is exclusively concerned with the systematic
error in estimating rates and proposes ``optimal'' methods that
minimize this systematic rate estimation error. Therefore all statements
are strictly valid only in the data-rich regime. Explicit treatment
of the statistical error in the data-poor regime is beyond the scope
of the present work, but is briefly discussed at the end of the paper 
and in the supplementary information.

\subsection*{Full-space dynamics}

We consider a dynamical system that follows a stationary and time-continuous 
Markov process $\mathbf{x}_{t}$ in its full (and generally large and continuous) 
phase space $\Omega$. 
 $\mathbf{x}_{t}$ is assumed to be ergodic
with a unique stationary density $\mu(\mathbf{x})$. In order to be
independent of specific dynamical models we use the general transition
density $p_{\tau}(\mathbf{x}_{t},\mathbf{x}_{t+\tau})$, i.e., the
conditional probability density that given the system is at point
$\mathbf{x}_{t}\in\Omega$ at time $t$, it will be found at point
$\mathbf{x}_{t+\tau}\in\Omega$ a lag time $\tau$ later. We will
at this point also assume that the dynamics obey microscopic detailed
balance, i.e.,
\begin{equation}
\mu(\mathbf{x}_{t})\, p_{\tau}(\mathbf{x}_{t},\mathbf{x}_{t+\tau})=\mu(\mathbf{x}_{t+\tau})\, p_{\tau}(\mathbf{x}_{t+\tau},\mathbf{x}_{t}),
\label{eq_detailed-balance-full}
\end{equation}
which is true for systems that are not driven by external forces.
In this case, $\mu(\mathbf{x})$ is a Boltzmann distribution in terms
of the system's Hamiltonian. 
In some dynamical models, e.g. Langevin dynamics, Eq. \ref{eq_detailed-balance-full}
does not hold, but rather some generalized form of it \cite{Barra_detailed_balance}. In this case, the present theory also
applies (see comment below), but in the interest of the simplicity of the equations, we will assume Eq. \ref{eq_detailed-balance-full}
subsequently.

For a two-state rate theory, we are interested in the slowest relaxation 
processes, and hence rewrite the transition density as a sum of relaxation 
processes (each associated with a different intrinsic rate) by expanding in 
terms of the eigenvalues $\lambda_i$ and eigenfunctions $\psi_{i}$ of the 
corresponding transfer operator
 \cite{PrinzEtAl_JCP10_MSM1,SarichNoeSchuette_MMS09_MSMerror}:
\begin{eqnarray}
	p_{\tau}(\mathbf{x}_{t},\mathbf{x}_{t+\tau}) & = & \sum_{i=1}^{\infty}\mathrm{e}^{-\kappa_{i}\tau}\psi_{i}(\mathbf{x}_{t})\mu(\mathbf{x}_{t+\tau})\psi_{i}(\mathbf{x}_{t+\tau}).\label{eq_spectral-decomposition}
\end{eqnarray}
Here, 
\begin{equation}
	\lambda_{i}(\tau)=e^{-\kappa_{i}\tau}\label{eq_propagator-eval}
\end{equation}
are eigenvalues of the propagator that decay exponentially with lag
time $\tau$. We order relaxation rates according to $\kappa_{1}<\kappa_{2}\le\kappa_{3}\le...$
and thus $\lambda_{1}(\tau)>\lambda_{2}(\tau)\ge\lambda_{3}(\tau)\ge...$.
The first term is special in that it is the only stationary process:
$\kappa_{1}=0$, $\lambda_{1}(\tau)=1$, $\psi_{1}(\mathbf{x})=1$,
thus the first term of the sum is identical to $\mu(\mathbf{x})$.
All other terms can be assigned a finite relaxation rate $\kappa_{i}$,
or a corresponding relaxation timescale $t_{i}=\kappa_{i}^{-1}$ ,
which are quantities of our interest. The eigenfunctions $\psi_{i}$
are independent of $\tau$ and determine the structure of the relaxation
process occurring with rate $\kappa_{i}$. The sign structure of $\psi_{i}(\mathbf{x})$
determines between which substates the corresponding relaxation process
is switching and is thus useful for identifying metastable sets, i.e.,
sets of states that are long-lived and interconvert only by rare
events \cite{SchuetteFischerHuisingaDeuflhard_JCompPhys151_146,PrinzEtAl_JCP10_MSM1}.
The eigenfunctions are chosen to obey the normalization conditions
\begin{equation}
\langle\psi_{i},\psi_{j}\rangle_{\mu}=\int_{\iO}d\mathbf{x}\:\psi_{i}(\mathbf{x})\psi_{j}(\mathbf{x})\mu(\mathbf{x})=\delta_{ij}\label{eq_normalization}
\end{equation}
and integration always runs over the full space of the integrated
variable if not indicated otherwise. At a given timescale $\tau$
of interest, fast processes with $\kappa\gg\tau^{-1}$ (and correspondingly
$t_{i}\ll\tau$) will have effectively vanished and we are typically
left with relatively few slowly-relaxing processes. 

Finally, we define the $\mu$-reweighted eigenfunctions,
\begin{equation}
\phi_{i}(\mathbf{x})=\mu(\mathbf{x})\psi_{i}(\mathbf{x})
\end{equation}
such that the normalization condition of eigenfunctions can be conveniently
written as
\begin{equation}
\langle\phi_{i},\psi_{j}\rangle=\int_{\iO}d\mathbf{x}\:\phi_{i}(\mathbf{x})\psi_{j}(\mathbf{x})=\delta_{ij}.
\end{equation}
Finally, the correlation density $c_{\tau}(\mathbf{x}_{t},\mathbf{x}_{t+\tau})$,
i.e., the joint probability density of finding the system at $\mathbf{x}_{t}$
at time $t$ and at $\mathbf{x}_{t+\tau}$ at time $t+\tau$ is related
to the transition density $p_{t}$ by
\begin{eqnarray}
c_{\tau}(\mathbf{x}_{t},\mathbf{x}_{t+\tau}) & = & \mu(\mathbf{x}_{t})\, p_{\tau}(\mathbf{x}_{t},\mathbf{x}_{t+\tau}).\label{eq_correlation-density}
\end{eqnarray}

\subsection*{Observed dynamics and  two-state spectral  rate theory}

Let us consider the case that we are only interested in a single relaxation
process --- the slowest. Below, we sketch a rate theory for this case.
Details of the derivation can be found in the supplementary information
(SI). Based on the definitions above, the correlation density can
then be written as:
\begin{eqnarray}
c_{\tau}(\mathbf{x}_{t},\mathbf{x}_{t+\tau}) & = & \mu(\mathbf{x}_{t})\mu(\mathbf{x}_{t+\tau})\nonumber \\
 &  & +\mathrm{e}^{-\kappa_{2}\tau}\mu(\mathbf{x}_{t})\psi_{2}(\mathbf{x}_{t})\mu(\mathbf{x}_{t+\tau})\psi_{2}(\mathbf{x}_{t+\tau})\nonumber \\
 &  & +\mathrm{e}^{-\kappa_{3}\tau}\mu(\mathbf{x}_{t})p_{\tau,\mathrm{fast}}(\mathbf{x}_{t},\mathbf{x}_{t+\tau})\label{eq_correlation-density-singlerelaxation}
\end{eqnarray}
where, if detailed balance (\ref{eq_detailed-balance-full}) holds, the fast processes are given by Eq. (\ref{eq_spectral-decomposition}):
\begin{equation}
p_{\tau,\mathrm{fast}}(\mathbf{x}_{t},\mathbf{x}_{t+\tau})=\sum_{i=3}^{\infty}\mathrm{e}^{-(\kappa_{i}-\kappa_{3})\tau}\psi_{i}(\mathbf{x}_{t})\mu(\mathbf{x}_{t+\tau})\psi_{i}(\mathbf{x}_{t+\tau})\label{eq:}
\end{equation}
If detailed balance does not hold on the full phase space, but rather some generalized form of it, the spectrum may have complex eigenvalues. Even in this case, the
fast part of the dynamics can be bounded by $\mathrm{e}^{-\kappa_{3}\tau}$, and therefore Eq. (\ref{eq_correlation-density-singlerelaxation}) and the subsequent theory holds. See also discussion in \cite{SarichNoeSchuette_MMS09_MSMerror}.

\emph{Exact rate}: 
$\kappa_2$ is often termed the \emph{phenomenological rate}, 
as it governs the dominant relaxation rate of any observed signal in which the slowest relaxation process is apparent.
The exact rate of interest $\kappa_{2}$ can be theoretically
recovered as follows: If we know the exact corresponding eigenfunction
$\psi_{2}(\mathbf{x})$, it follows from Eq.~(\ref{eq_spectral-decomposition})
and (\ref{eq_normalization}) that its autocorrelation function evaluates
to:
\begin{eqnarray}
\lambda_{2}(\tau) & = & \langle\psi_{2}(\mathbf{x}_{t})\psi_{2}(\mathbf{x}_{t+\tau})\rangle_{t}\nonumber \\
 & = & \int_{\iO}d\mathbf{x}_{t}\,\int_{\iO}d\mathbf{x}_{t+\tau}\, c_{\tau}(\mathbf{x}_{t},\mathbf{x}_{t+\tau})\,\psi_{2}(\mathbf{x}_{t})\,\psi_{2}(\mathbf{x}_{t+\tau})\nonumber \\
 & = & e^{-\kappa_{2}\tau}\label{eq_exact-rate}
\end{eqnarray}
where $\langle\cdot\rangle_{t}$ denotes the time average, that is
here identical to the ensemble average due to ergodicity of the dynamics.

The correlation function $\langle\psi_{2}(0)\psi_{2}(\tau)\rangle_{t}$ yields the exact eigenvalue
$\lambda_{2}(\tau)$ and thus also an exact rate estimate $\hat{\kappa}_{2}=-\tau^{-1}\ln\lambda_{2}(\tau)=\kappa_{2}$,
independently of the choice of $\tau$. 

\begin{figure*}
\begin{centering}
	\includegraphics[width=0.85\textwidth]{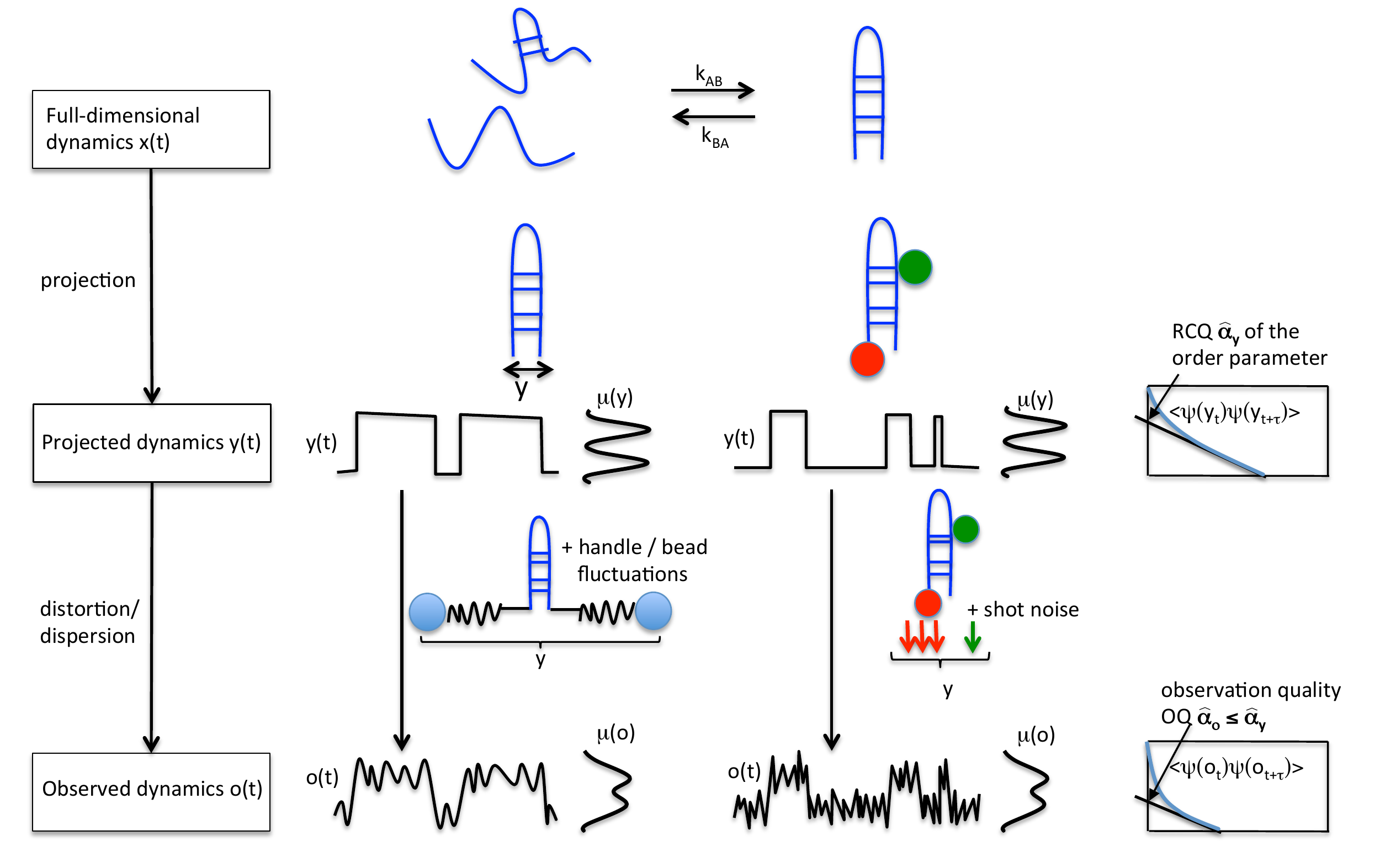}
\par\end{centering}

\caption{\label{fig_scheme-observation}
{\bf Illustration of the observed dynamics for which a rate theory is formulated here.}
\emph{Top row:} the full-dimensional
dynamics $\mathbf{x}(t)$ in phase space $\Omega$. These dynamics
are assumed to be Markovian, ergodic, and reversible as often found
for physical systems in thermal equilibrium. Furthermore, the theory
here is formulated for two-state kinetics, i.e., the system has two
metastable states exchanging at rates $k_{AB}$ and $k_{BA}$, giving
rise to a relaxation rate of $\kappa_{2}=k_{AB}+k_{BA}$. \emph{Middle
row:} One order parameter $y(\mathbf{x})$ of the system is observed,
such as a distance between two groups of a molecule, or the FRET efficiency
between two fluorescent groups. The projection of the full-space dynamics
$\mathbf{x}(t)$ onto the order parameter $y$ generates a time series
$y(t)$, that may, however not be directly observable. The projection
also acts on functions of state space, such as the stationary distribution
of configurations in full state space that is projected onto a density
in the observable, $\mu(y)$. The reaction coordinate quality $\hat{\alpha}_{y}$
measures how well the order parameter $y$ resolves the slow transition.
It is 1 when $A$ and $B$ are perfectly separated and $0$ when they
completely overlap. \emph{Bottom row:} the experimental device used
may distort or disperse the signal, for example by adding noise. The
resulting observed signal $o(t)$ is distorted and the observable
density $\mu^{o}(o)$ is smoothed. $\hat{\alpha}_{o}$ measures the
observation quality (OQ) of the observed signal and it is shown
in the SI that $\hat{\alpha}_{o}\le\hat{\alpha}_{y}$ holds.}
\end{figure*}

\emph{Projected} \emph{dynamics without measurement noise}: Suppose
we observe the dynamics of an order parameter $y\in\mathbb{R}$ that
is a function of the configuration $\mathbf{x}$. Examples are the
distance between two groups of the molecule, or a more complex observable
such as the F\"orster resonance transfer efficiency associated to a
given configuration. See Fig.~\ref{fig_scheme-observation} for an
illustration. We first assume that no additional measurement noise
is present. The analysis of a molecular dynamics simulation where
a given order parameter is monitored is one example of such a scenario.
Now, it is no longer possible to compute the rate via Eq.~(\ref{eq_exact-rate})
or some direct approximation of Eq.~(\ref{eq_exact-rate}), since
the full configuration space $\Omega$ in which the eigenfunction
$\psi_{2}$ exists can no longer be recovered once the dynamics has
been projected onto an order parameter. Instead, we are forced to
work with functions of the observable $y$. While the theory is valid for 
multidimensional observables $y$, the equations below assume $y\in\mathbb{R}$ for simplicity.

We have two options for deriving the relevant rate equations for the
present scenario: As a first option, we note that a projection that
is free of noise can be regarded as a function $y(\mathbf{x}):\Omega\rightarrow\mathbb{R}$.
Thus, any function $\tilde{\psi}_{2}(y)$ of elements in observable
space $\mathbb{R}$ that aims at approximating the dominant eigenfunction
$\psi_{2}$, can be also regarded as a function in full space $\Omega$
via $\tilde{\psi}_{2}(y)=\tilde{\psi}_{2}(y(\mathbf{x}))$. When following
this idea, one can use the variational principle of conformation dynamics
\cite{Noe:2012ve} (see also the discrete-state treatment in \cite{Buchner_BBA11_ProteinFoldingKinetics}),
in order to derive the rate equations for the observed space dynamics.
See SI for details.

However, since we aim  to include the possibility of  measurement noise in a second step,
we derive a more general approach (see SI), which is summarized subsequently.
Consider the function $\chi_{p}(y\mid\mathbf{x})$ that denotes the
output probability density with which each configuration of the full
state space, $\mathbf{x}\in\Omega$, yields a measured value $y\in\mathbb{R}$.
In the case of simply projecting $\mathbf{x}$-values without noise
to specific $y$-values, $\chi$ has the simple form:
\begin{equation}
\chi_{p}(y'\mid\mathbf{x})=\delta(y'-y(\mathbf{x}))\label{eq_chi_p}
\end{equation}
This allows the correlation density in the observable space to be written
as:
\begin{eqnarray}
 &  & c_{\tau}(y_{0},y_{\tau})\nonumber \\
 & = & \int_{\iO}d\mathbf{x}_{0}\,\int_{\iO}d\mathbf{x}_{\tau}\,\chi_{p}(y_{0}\mid\mathbf{x}_{0})c_{\tau}(\mathbf{x}_{0},\mathbf{x}_{\tau})\chi_{p}(y_{\tau}\mid\mathbf{x}_{\tau})\nonumber \\
 & = & \mu^{y}(y_{0})\mu^{y}(y_{\tau})+\sum_{i=2}^{\infty}\lambda_{i}(\tau)\phi_{i}^{y}(y_{0})\phi_{i}^{y}(y_{\tau})\label{eq_corr-y}
\end{eqnarray}
where we have used superscript $y$ to indicate the projection of
a full configuration space function onto the order parameter: $\mu^{y}(y)$
is the observed stationary density that can be estimated from a sufficiently
long simulation by histogramming the values of $y$. Mathematically,
it is given by:
\begin{equation}
\mu^{y}(y)=\int_{\iO}d\mathbf{x}\:\chi_{p}(y\mid\mathbf{x})\mu(\mathbf{x})\label{eq_statdist_y}
\end{equation}
$\phi_{i}^{y}$ are the projected eigenfunctions:
\begin{equation}
\phi_{i}^{y}(y)=\int_{\iO}d\mathbf{x}\:\chi_{p}(y\mid\mathbf{x})\phi_{i}(\mathbf{x})\label{eq_eigenfun_y}
\end{equation}
In order to arrive at an expression of the rate $\kappa_{2}$, we
propose a trial function in observation space, $\tilde{\psi}_{2}(y)$,
which we require to be normalized by
\begin{eqnarray}
	\langle\tilde{\psi}_{2},1\rangle_{\mu^{y}}=0, & \:\:\:\: & \langle\tilde{\psi}_{2},\tilde{\psi}_{2}\rangle_{\mu^{y}}=1,\label{eq_psi2_observable_normalization-1}
\end{eqnarray}
and evaluate its autocorrelation function as: 
\begin{eqnarray}
	\langle\tilde{\psi}_{2}(y_{0})\tilde{\psi}_{2}(y_{\tau})\rangle & = & \int_{\iR}dy_{0}\int_{\iR}dy_{\tau}\,\tilde{\psi}_{2}(y_{0})c_{\tau}(y_{0},y_{\tau})\tilde{\psi}_{2}(y_{\tau})\nonumber \\
 & = & \alpha_{y}\mathrm{e}^{-\kappa_{2}\tau}+\sum_{i>2}\langle\tilde{\psi}_{2},\phi_{i}^{y}\rangle^{2}\mathrm{e}^{-\kappa_{i}\tau}
\end{eqnarray}
where
\begin{equation}
	\alpha_{y}=\langle\tilde{\psi}_{2},\phi_{2}^{y}\rangle^{2}.
	\label{eq_alpha_y}
\end{equation}
In contrast to Eq.~(\ref{eq_exact-rate}), both $\tilde{\psi}_{2}$
and $\phi_{i}^{y}$ live on the observable space $\mathbb{R}$. In the special case
that $\psi_{2}(\mathbf{x})$ is constant in all other variables than
$y(\mathbf{x})$, the projection is lossless ($\phi_{2}^{y}(y(\mathbf{x}))=\phi_{2}(\mathbf{x})$
and $\psi_{2}^{y}(y(\mathbf{x}))=\psi_{2}(\mathbf{x})$ for all $\mathbf{x}$),
and using the choice $\tilde{\psi}_{2}=\psi_{2}^{y}$, we recover
$\langle\tilde{\psi}_{2}^{y},\phi_{2}^{y}\rangle=\langle\psi_{2},\phi_{2}\rangle=1$,
and thus the exact rate estimate via Eq.~(\ref{eq_exact-rate}).
In general, however, the eigenfunction $\psi_{2}(\mathbf{x})$ does
vary in other variables than $y$, and therefore $\tilde{\psi}_{2}$
can at best approximate the full-space eigenfunction via $\tilde{\psi}_{2}(y(\mathbf{x}))\approx\psi_{2}(\mathbf{x})$. 

\emph{Observed dynamics with measurement noise}: Suppose that an experiment
is conducted in which each actual order parameter value $y(\mathbf{x})\in\mathbb{R}$
is measured with additional noise, yielding the observed value $o\in\mathbb{R}$.
In time-binned single-molecule fluorescence experiments such noise
may come from photon counting shot noise for a given binning size.
In optical tweezer experiments such noise may come from bead diffusion
and handle elasticity, assuming that bead and handle dynamics are
faster than the kinetics of the molecule of interest. See Fig.~\ref{fig_scheme-observation}
for an illustration. Note that we only treat the situation of uncorrelated
noise. In situations where the experimental  configuration  changes the kinetics,
e.g., when the optical bead diffusion is slow, thus exhibiting different
transition rates than the isolated molecule, our analysis always reports
the rate of the  overall observed system. The task of correcting the measured
rates so as to estimate the rates of the pure molecule is beyond the
scope of this work and can, for example, be attempted via dynamical
deconvolution \cite{HinczewskiVonHansenNetz_PNAS10_DynamicalDeconvolution,vonHansenNetz_RevSciInstr12_SpectralAnalysisTweezer} or other approaches 
\cite{Thirumalai_PNAS2013}. % http://dx.doi.org/10.1073/pnas.1214051110

Like before, the probability of observing a measurement value $o\in\mathbb{R}$
given that the true configuration was $\mathbf{x}\in\Omega$ can be
given by an output probability:
\begin{equation}
\chi_{pd}(o\mid\mathbf{x})=\int_{\iR}dy\,\chi_{d}(o\mid y)\chi_{p}(y\mid\mathbf{x}),
\end{equation}
which  convolves  the projection from $\mathbf{x}$ to the value
of the order parameter, $\chi_{p}(y\mid\mathbf{x})$,  with  the subsequent
dispersion of the signal by noise, $\chi_{d}(o\mid y)$. Despite the
fact that dispersion operates by a different physical process than
projection, the same analysis as above applies. We define the projected
and dispersed stationary density and eigenfunctions:
\begin{eqnarray}
\mu^{o}(o) & = & \int_{\iO}d\mathbf{x}\:\chi_{pd}(o\mid\mathbf{x})\mu(\mathbf{x}) \nonumber \\ 
	& = & \int_{\iR}dy\:\chi_{d}(o\mid y)\mu^{y}(y)\\
\phi_{i}^{o}(o) & = & \int_{\iO}d\mathbf{x}\:\chi_{pd}(o\mid\mathbf{x})\phi_{i}(\mathbf{x}) \nonumber \\ 
	& = & \int_{\iR}dy\:\chi_{d}(o\mid y)\phi_{i}^{y}(y)
\end{eqnarray}
which are ``smeared out'' by noise compared to the purely projected
density and eigenfunctions $\phi_{i}^{y}$. As above, the autocorrelation
function of a probe function $\tilde{\psi}_{2}(o)$ is given by: 
\begin{eqnarray}
\langle\tilde{\psi}_{2}(o_{0})\tilde{\psi}_{2}(o_{\tau})\rangle & = & \alpha_{o}\mathrm{e}^{-\kappa_{2}\tau}+\sum_{i>2}\langle\tilde{\psi}_{2},\phi_{i}^{o}\rangle^{2}\mathrm{e}^{-\kappa_{i}\tau}
\end{eqnarray}
with
\begin{equation}
\alpha_{o}=\langle\tilde{\psi}_{2},\phi_{2}^{o}\rangle^{2}.
\end{equation}
The observation process including noise is a more general process
than the observation process  excluding  noise, therefore --- unless
the distinction is important --- we will generally refer to the observation
as $o$ subsequently, whether noise is included in the observation or not.

\emph{Filtered dynamics}: The effect of measurement noise may be reduced
by filtering  (averaging)  the observed signal $o(t)\rightarrow\bar{o}(t)$, for
example by averaging the signal value over a time window of length
$W$. Note that this operation will introduce memory of length $W$
into the signal and will impair the estimation of all rates which
are close to $W^{-1}$. Fig.~1 of the SI illustrates the effect of
filtering on the estimation quality of rates in a simple example.
To make sure that the filter used does not impair the rate estimates,
we recommend that the filter length be at least a factor of 10 smaller
than the timescales of interest, $t_{2}=\kappa_{2}^{-1}$. The filtered
signal $\bar{o}(t)$ can then be used as input to the various rate
estimators discussed in this paper, but the theory of systematic errors
given in the subsequent section may no longer apply because filtering
destroys the Markovianity of the original dynamic process in the full
state space.
A more extensive treatment of filtering is given in the SI.

\emph{Direct rate estimate}: In all of the above cases, the autocorrelation
function of the trial function $\tilde{\psi}_{2}$ does not yield
the exact eigenvalue $\lambda_{2}(\tau)$, but some approximation
$\tilde{\lambda}_{2}(\tau)$. For $\tau\gg\kappa_{3}^{-1}$, which
can readily be achieved for clear two-state processes where a time
scale separation exists ($\kappa_{2}\gg\kappa_{3}$), the terms involving
the fast processes disappear:
\begin{equation}
\tilde{\lambda}_{2}(\tau)\approx\alpha_{o}e^{-\kappa_{2}\tau}.\label{eq_eval-estimate}
\end{equation}
This suggests that the true rate $\kappa_{2}$, as well as the prefactor
$\alpha_{o}$ that may serve as a basis to measure the observation quality, 
could be recovered from large $\tau$ decay of an appropriately
good trial function even from the observed signal. We elaborate this
concept in subsequent sections. Note that in experiments the relaxation
rates $\kappa_{2},\kappa_{3},$ etc, are initially unknown and hence
the validity of Eq.~(\ref{eq_eval-estimate}) can only be checked
a posteriori, e.g., by the fact that estimates based upon Eq.~(\ref{eq_eval-estimate})
are independent of the lag time $\tau$.

\subsection*{Existing rate estimators}

Many commonly used rate estimators consist of two steps: (1) they
(explicitly or implicitly) calculate an autocorrelation function $\tilde{\lambda}_{2}(\tau)$
of some function $\tilde{\psi}_{2}$, and (2) transform $\tilde{\lambda}_{2}(\tau)$
into a rate estimate $\tilde{\kappa}_{2}$. In order to derive an
optimal estimator, it is important to understand how the systematic
error of the estimated rate depends on each of the two steps. 
Therefore, we now rephrase existing rate estimators in the formalism of spectral rate theory.
The SI contains a detailed derivation of the subsequent results.

Many rate estimators operate by defining a single dividing surface
which splits the state space into reactants $A$ and products $B$.
Calling $h_{A}(o)$ the indicator function which is 1 for set $A$
and 0 for set $B$, one may define the normalized fluctuation autocorrelation
function of state $A$ \cite{Onsager_PhysRev1931_NumberCorrelationFunction}
\begin{equation}
\tilde{\lambda}_{2}(t)=\frac{\langle h_{A}(0)h_{A}(\tau)\rangle-\langle h_{A}\rangle^{2}}{\left\langle h_{A}^{2}\right\rangle -\langle h_{A}\rangle^{2}}=\langle\tilde{\psi}_{2}(0)\tilde{\psi}_{2}(\tau)\rangle\label{eq_number-correlation-function-1-1}
\end{equation}
that can also be interpreted as an autocorrelation function $\tilde{\lambda}_{2}(t)$
with a step function $\tilde{\psi}_{2,\mathrm{divide}}(o)=(h_{A}(o)-\pi_{A})/\sqrt{\pi_{A}\pi_{B}}$.
Here, $\pi_{A}=\langle h_{A}\rangle_{\mu}$ is the stationary probability
of state $A$ and $\pi_{B}=1-\pi_{A}$ the stationary probability
of state $B$. Other rate estimates choose $\tilde{\psi}_{2}$ to
be the signal $o(t)$ itself or the committor function between two
pre-defined subsets of the $o$ coordinate \cite{SchuetteEtAl_JCP11_Milestoning}.
We show that none of these choices is optimal, and the optimal choice
of $\tilde{\psi}_{2}$ will be derived in the subsequent section.

Existing rate estimators largely differ by step (2), i.e., how they
transform $\tilde{\lambda}_{2}(t)$ into a rate estimate $\tilde{\kappa}_{2}$.
This procedure then determines the functional form of the systematic
estimation error. We subsequently list bounds for these errors (see
SI for derivation). The prefactor $\alpha$ in the equations below
either refers to $\alpha_{p}$ (purely projected dynamics) or $\alpha_{o}$
(dynamics with noise), whichever is appropriate. 

\emph{Reactive flux rate.} Chandler, Montgomery and Berne~\cite{Chandler_JCP78_ReactiveFlux,MontgomeryChandlerBerne_JCP79}
considered the reactive flux correlation function as a rate estimator:
$\tilde{\kappa}_{2,\mathrm{rf}}(\tau)=-\frac{d}{dt}\tilde{\lambda}_{2}(\tau)$.
Its error is
\begin{eqnarray}
\tilde{\kappa}_{2,\mathrm{rf}}-\kappa_{2} & = & \kappa_{2}(\alpha-1)+\sum_{i>2}\langle\tilde{\psi}_{2},\psi_{i}\rangle_{\mu}^{2}\kappa_{2}e^{-\kappa_{i}\tau}>0\nonumber \\
\label{eq:-3}
\end{eqnarray}
which becomes 0 for the perfect choice of $\tilde{\psi}_{2}=\psi_{2}$ that leads to $\alpha=1$,
but can be very large otherwise.

\emph{Transition state theory rate.} 
The transition state theory rate, which measures the instantaneous flux across the dividing surface
between $A$ and $B$, is often estimated by the trajectory length divided by the number of crossings of the dividing surface. Its simplicity makes 
it a widely popular choice for practical use in experiments and theory (despite its tendency to produce biased estimates, as we will discuss later).

In order to arrive at an expression for the estimation error, the TST rate can be expressed as the short-time limit of
the reactive flux~\cite{Chandler_JCP78_ReactiveFlux}, $\hat{\kappa}_{2,\mathrm{tst}}=\lim_{\tau\rightarrow0^{+}}\tilde{\kappa}_{2,\mathrm{rf}}(\tau)$
such that the error in the rate is given by
\begin{eqnarray}
\tilde{\kappa}_{2,\mathrm{tst}}-\kappa_{2} & = & \kappa_{2}(\alpha-1)+\sum_{i>2}\langle\tilde{\psi}_{2},\psi_{i}\rangle_{\mu}^{2}\kappa_{2}>\tilde{\kappa}_{2,\mathrm{rf}}-\kappa_{2}\nonumber \\
\label{eq:-4}
\end{eqnarray}
which is always an overestimate of the true rate and of the reactive flux rate. 

\emph{Integrating the correlation function.} Another means of estimating
the rate is via the integral of the correlation function, $\tilde{\kappa}_{2,\mathrm{int}}=-\left(\int_{0}^{\infty}d\tau\:\tilde{\lambda}_{2}(\tau)\right)^{-1}$
(see, e.g., Eq.~3.6 of \cite{Chandler_JCP78_ReactiveFlux}),
with the error:
\begin{eqnarray}
\tilde{\kappa}_{2,\mathrm{int}}-\kappa_{2} & = & \kappa_{2}\left(\frac{1-\alpha+\sum_{i>2}\langle\psi_{i},\tilde{\psi}_{2}\rangle_{\mu}^{2}\frac{\kappa_{2}}{\kappa_{i}}}{\alpha+\sum_{i>2}\langle\psi_{i},\tilde{\psi}_{2}\rangle_{\mu}^{2}\frac{\kappa_{2}}{\kappa_{i}}}\right)
\end{eqnarray}
in the special case that $\kappa_{3}\gg\kappa_{2}$ (time scale separation),
the error is approximately given by $\kappa_{2}(1-\alpha)/\alpha$.
Thus, the error of this estimator becomes zero for $\alpha=1$, which is only the case for a reaction coordinate with no noise, and no
further projection (e.g. by using a dividing surface). The error may be very large in other cases ($\alpha<1$).

\emph{Single-$\tau$ rate estimators}: A simple rate estimator takes the value of the autocorrelation function of some function $\tilde{\psi}_{2}$ at a single value of $\tau$, and transforms it into a rate estimate by virtue of Eq.~(\ref{eq_eval-estimate}).

We call these estimators \emph{single-$\tau$ estimators}. Ignoring
statistical uncertainties, they yield a rate estimate of the form
\begin{eqnarray}
\tilde{\kappa}_{2,\mathrm{single}} & = & -\frac{\ln\tilde{\lambda}_{2}(\tau)}{\tau}\label{eq_rate-estimate}
\end{eqnarray}
Quantitatively, the error can be bounded by the expression (see derivation
in the SI):
\begin{eqnarray}
\tilde{\kappa}_{2,\mathrm{single}}-\kappa_{2} & \le & -\frac{\ln\alpha}{\tau}.\label{eq_rate-error}
\end{eqnarray}
The error becomes identical to this bound for systems with a strong
timescale separation, $\kappa_{3}\gg\kappa_{2}$. Eq.~(\ref{eq_rate-error})
decays relatively slowly in time (with $\tau^{-1}$. See \refFig{fig:Rate-Estimate}
for a two-state example). It will be shown below that methods that
estimate rates from counting the number of transitions across a dividing
surface, such as Markov state models (MSMs), are single-$\tau$ estimators
and are thus subject to the error given by Eq.~(\ref{eq_rate-error}). 

The systematic error of single-$\tau$ estimators results from the
fact that Eq.~(\ref{eq_rate-estimate}) effectively attempts to fit
the tail of a multi-exponential decay $\tilde{\lambda}_{2}(\tau)$
by a single-exponential with the constraint $\tilde{\lambda}_{2}(0)=1$.
Unfortunately, the ability to improve these estimators by simply increasing
$\tau$ is limited because the statistical uncertainty of estimating
Eq.~\ref{eq_eval-estimate} quickly grows with increasing  $\tau$ \cite{ZwanzigAilawadi_PRE69_StatisticalErrorCorrelations}. 

\emph{Multi-$\tau$ rate estimators}: To avoid the error given by
Eq.~(\ref{eq_rate-error}), it is advisable to estimate the rate
by evaluating the autocorrelation function $\tilde{\lambda}_{2}(\tau)$
at multiple values of $\tau$. This can be done e.g., by performing
an exponential fit to the \emph{tail} of the $\tilde{\lambda}_{2}(\tau)$,
thus avoiding the constraint $\tilde{\lambda}_{2}(0)=1$ \cite{SkinnerWolynes_JCP78_RelaxationProcesses,Zhou_QuartRevBiophys10_RateReview}.
The corresponding estimation error $\hat{\kappa}_{2,\mathrm{mult}}-\kappa_{2}$ is bounded by:
\begin{equation}
\hat{\kappa}_{2,\mathrm{mult}}-\kappa_{2}<c\frac{1-\alpha}{\alpha}\mathrm{e}^{-\tau_{1}(\kappa_{3}-\kappa_{2})}\label{eq_multi-tau-error}
\end{equation}
where $\tau_{1}$ is the first lag time from the series $(\tau_{1},...,\tau_{m})$
used for fitting, and the constant $c$ also depends on the lag times
and the fitting algorithm used. The SI shows that for several fitting
algorithms, such as a least-squares procedure at the time points $(\tau,\,2\tau,\,...,\, m\tau)$,
$c$ is such that
\begin{equation}
\hat{\kappa}_{2,\mathrm{mult}}\le\tilde{\kappa}_{2,\mathrm{single}}.\label{eq_mult-is-better-than-single}
\end{equation}
Thus, the multi-$\tau$ estimator is always better than the single-$\tau$
estimator (see SI). The main advantage of multi-$\tau$ estimators
is that their convergence rate is exponential in $\tau$ when the
time scale separation $\kappa_{3}-\kappa_{2}$ is not vanishing (compare
to Eq.~\ref{eq_rate-error}). Thus, multi-$\tau$ estimators are
better when the timescale separation between the slowest and the other
relaxation rates in the system is larger. 
%Note that the systematic
%error of a multi-$\tau$ estimate thus decays much faster in $\tau$
%than its statistical error that decays by $\tau^{-1}$ \cite{ZwanzigAilawadi_PRE69_StatisticalErrorCorrelations}. 

In the absence of statistical error, all of the above rate estimation
methods  are seen to  yield an overestimation of the rate, $\tilde{\kappa}_{2}\ge\kappa_{2}$.

\subsection*{Optimal choice of $\tilde{\psi}_{2}$}

It was shown above that multi-$\tau$ estimators are the best choice
for converting an autocorrelation function into a rate estimate. However,
what is the best possible choice $\hat{\text{\ensuremath{\psi}}}_{2}=\tilde{\psi}_{2,\mathrm{optimal}}$
given a specific observed time series $o_{t}$? In other words, which
function should the observed dynamics be projected upon in order to
obtain an optimal rate estimator? Following Eq.~(\ref{eq_rate-error}),
the optimal choice $\hat{\text{\ensuremath{\psi}}}_{2}$ is the one
which maximizes the parameter $\alpha$, as this will minimize the
systematic error from a direct rate estimation by virtue of Eq.~(\ref{eq_rate-error}),
and also minimize the systematic error involved in estimating $\kappa_{2}$
from an exponential fit to Eq.~(\ref{eq_eval-estimate}). We are
thus seeking the solution of:
\begin{eqnarray}
\hat{\psi}_{2} & = & \arg\max_{\tilde{\psi}_{2}}\alpha=\arg\max_{\tilde{\psi}_{2}}\tilde{\lambda}_{2}(\tau)\label{eq_alpha-optimization}
\end{eqnarray}
 for some $\tau > 0$, 
subject to the normalization Eq.~(\ref{eq_psi2_observable_normalization-1}).
Here, $\arg\max_{\tilde{\psi}_{2}}\alpha$ denotes the function that
maximizes $\alpha$ over the space of functions $\tilde{\psi}_{2}(o)$.
If the system has two-state kinetics, i.e., only $\psi_{1}(\mathbf{x})=1$
and $\psi_{2}(\mathbf{x})$ are present as dominant eigenfunctions,
the problem (\ref{eq_alpha-optimization}) is solved by the projected
eigenfunction:
\begin{equation}
\hat{\psi}_{2}=\psi_{2}^{o}
\end{equation}
How can the best possible $\hat{\psi}_{2}$ be determined from the
observed time series? For a sufficiently large set of $n$ basis functions,
$\boldsymbol{\gamma}=\{\gamma_{1}(o),...,\gamma_{n}(o)\}$, the optimal
eigenfunction $\hat{\psi}_{2}$ is approximated by a linear combination
$\hat{\psi}_{2}(o)\approx\sum_{i=1}^{n}c_{i}\gamma_{i}(o)$ with coefficients
$\mathbf{c}=\{c_{1},...,c_{n}\}$. When $\boldsymbol{\gamma}$ is chosen
to be an orthogonal basis set, then $\hat{\psi}_{2}=\arg\max_{\tilde{\psi}_{2}}\alpha$
can be approximated by the Ritz method \cite{Ritz09-Variationsproblem,Noe:2012ve}.
An easy way to do this approximation in practice is to perform a fine
discretization of the observable $o$ by histogram windows. Using
a binning with bin boundaries $b_{1},...,b_{n+1}$, and the corresponding
indicator functions 
\begin{equation}
\gamma_{i}(o)=\begin{cases}
1 & \text{if }i\in[b_{i},b_{i+1})\\
0 & \text{else}
\end{cases},
\end{equation}
then the above optimization problem is solved by estimating the transition
probability matrix with elements
\begin{equation}
T_{ij}=\mathbb{P}\left(o(\tau)\in[b_{j},b_{j+1})\mid o(0)\in[b_{i},b_{i+1})\right)\label{eq_transition-matrix-binned}
\end{equation}
and calculating $\mathbf{c}$ as the second eigenvector: 
\begin{equation}
\mathbf{T}\mathbf{c}=\lambda_{2}\mathbf{c}\label{eq_1d-transition-matrix}
\end{equation}
where $\lambda_{2}<1$ is the second-largest eigenvalue of $\mathbf{T}$.
If the system has two-state kinetics, i.e., only $\psi_{1}(\mathbf{x})=1$
and $\psi_{2}(\mathbf{x})$ are present as dominant eigenfunctions,
the estimate $\hat{\psi}_{2}$ is independent of the choice of $\tau$
in Eq.~(\ref{eq_transition-matrix-binned}). Thus, in real systems,
$\tau$ should be chosen to be at least a multiple of $\kappa_{3}^{-1}$
(e.g., $\tau\ge3\kappa_{3}^{-1}$, as indicated by a constant rate
$\kappa_{2}$ estimate using a multi-$\tau$ estimator (Eq.~(\ref{eq_multi-tau-error})).
Note that a given optimal $\hat{\psi}_{2}(o)$ can still be used with
single-$\tau$ and multi-$\tau$ rate estimators that would produce
different estimates for $\kappa_{2}$.

Note that $\hat{\psi}_{2}$ according to the procedure described here
is only optimal for the case when the observed signal is obtained
by projecting the high-dimensional data onto the observable, but is
no longer optimal in the presence of noise, and especially large noise.
In order to choose $\hat{\psi}_{2}$ optimal when noise is present,
a generalized Hermitian eigenvalue problem must be solved instead
of Eq.~(\ref{eq_1d-transition-matrix}) which includes a mixing matrix
whose elements quantify how much the observable bins are mixed due
to measurement noise. Since this approach is not very straightforward
and in most practical cases only leads to small improvements,  we do not pursue this approach further here.  
We rather note that it is often practical to reduce
the noise level by carefully filtering the recorded data, provided
that the filter length is much shorter than the timescales of interest.

\subsection*{Reaction coordinate quality (RCQ), estimation quality, and observation quality (OQ)}

Evaluating the suitability of a given observable for capturing the
slow kinetics is of great general interest. Although there is not
a unique way of quantifying this suitability of the observable, the
term reaction coordinate quality (RCQ) is often used. Previous studies
have proposed ways to measure the RCQ that are based on comparing
the observed dynamics to specific dynamical models or testing the
ability of the observable to model the committor or splitting probability
between two chosen end-states $A$ and $B$ \cite{Peters_jcp06_committorerrorestimation,MorrisonThirumalai_PRL11_SingleMoleculePulling,DudkoGrahamBest_PRL2011_FoldingBarrier}.
These metrics are either only valid for specific models of dynamics
or themselves require a sufficiently good separation of $A$ and $B$
by definition, restricting their applicability to observables with
rather good RCQs. 

The prefactor $\hat{\alpha}_{y}$ (see also Fig. \ref{fig_scheme-observation}  is a
measure between 0 and 1, quantifying the relative amplitude of the
slowest relaxation in the autocorrelation function after projection of the full-space
dynamics onto the molecular observable employed. The value $\hat{\alpha}_{y}$ only depends on the observable itself,
and is free of modeling choices and of the way rates are estimated from the signal. 
Therefore, we propose that $\alpha_{y}$ \textit{is} the reaction coordinate quality (RCQ).

However, $\alpha_{y}$ is not directly measurable: for a given observation, both the projection of the full-space dynamics
and measurement noise compromise the quality of the signal, and these effects cannot be easily separated. 
In addition, the actual pre-factor that is obtained in a given estimate of the signal autocorrelation function, $\alpha_{o}$
depends on the way the data is analyzed, namely the functional form $\tilde{\psi}_{2}(o)$ used to
compute the autocorrelation function $\tilde{\lambda}_{2}(\tau)$. Therefore, $\alpha_{o}$ is just an estimation quality.

Fortunately, the ambiguity of the estimation quality is removed for the optimal choice
$\tilde{\psi}_{2}=\hat{\psi}_{2}$ (Eq.~(\ref{eq_alpha-optimization})),
which maximizes $\alpha_{o}$. In this case we denote this prefactor
$\hat{\alpha}_{o}$, where $\hat{\alpha}_{o}=\alpha_{o}(\hat{\psi}_{2})\ge\alpha_{o}(\tilde{\psi}_{2})$.
Since $\hat{\alpha}_{o}$ only depends on the observed signal, and not on the way of analyzing it, we term it
observation quality (OQ). The OQ is a very important quantity, because by virtue of Eqs.~(\ref{eq_rate-error})
and (\ref{eq_multi-tau-error}), $\hat{\alpha}_{o}$ quantifies how large the error in our rate estimate can be
for the optimal choice $\tilde{\psi}_{2}=\hat{\psi}_{2}$.

Our definitions of RCQ and OQ are very general,
as they makes no assumptions about the class of dynamics in the observed
coordinate, and does not depend on any subjective choices such as
the choice of two reaction end-states $A$ and $B$ in terms of the
observable $o$. Through the derivation above it has also been shown
that $\hat{\alpha}_{o}$ measures the fraction of amplitude by which
the slowest process is observable, which is exactly the property one
would expect from a measure of the RCQ: $\hat{\alpha}_{0}$ is 1 for
a perfect reaction coordinate with no noise, and 0 if the slowest process is exactly
orthogonal to the observable, or has been completely obfuscated by
noise.

While the OQ is the quantity that can be computed from the signal, an analyst is typically
interested in the RCQ $\hat{\alpha}_{y}$ that is due to the choice
of the molecular order parameter. Unless a quantitative model of the dispersion function
$\chi_{d}(o\mid y)$ is known, the RCQ $\hat{\alpha}_{y}$ before
adding noise cannot be recovered. (see also Fig.~\ref{fig_scheme-observation}
for an illustration). However, we can still quantitatively relate
$\hat{\alpha}_{y}$ and $\hat{\alpha}_{o}$, and thereby show that even the OQ is very useful. 
For this, we derive a theory of observation quality. While the detailed derivation is found
in the SI, we just summarize the most important results here: 
\begin{enumerate}
\item When observing the order parameter $y$ without noise and projecting
the observation onto the optimal indicator function $\tilde{\psi}_{2}=\hat{\psi}_{2}^{o}$,
the RCQ can be expressed as the weighted norm of the projected eigenfunction,
expressed by the scalar product: 
\begin{equation}
\hat{\alpha}_{y}=\langle\phi_{2}^{y},\psi_{2}^{y}\rangle
\end{equation}

\item Unless the projection perfectly preserves the structure of the full-space
eigenfunction $\psi_{2}$, we have $\hat{\alpha}_{y}<1$. Thus, almost
every observable reduces the RCQ.
\item When additional noise is present, the OQ can be expressed as the
weighted norm of the projected and noise-distorted eigenfunction:
\begin{equation}
\hat{\alpha}_{o}=\langle\phi_{2}^{o},\psi_{2}^{o}\rangle
\end{equation}

\item The RCQ $\hat{\alpha}_{y}$ due to the projection onto the selected
molecular order parameter alone, and the OQ $\hat{\alpha}_{o}$
including measurement noise are related by: 
\begin{equation}
\hat{\alpha}_{o}\leq\hat{\alpha}_{y},\label{eq_RCQ-ordering}
\end{equation}
i.e., adding noise means that the OQ is smaller than the RCQ.
\end{enumerate}
The inequality (\ref{eq_RCQ-ordering}) implies that we can use the
OQ $\hat{\alpha}_{o}$ in order to both optimize the experimental
setup and the order parameter used. For example in an optical tweezer
measurement, we can change laser power and handle length so as to
maximize $\hat{\alpha}_{o}$, thus making $\hat{\alpha}_{o}$ and
$\hat{\alpha}_{y}$ more similar and reducing the effect of noise
on the measurement quality. On the other hand, since $\hat{\alpha}_{o}$
is a lower bound for $\hat{\alpha}_{y}$, we can also use it ensure
a minimal projection quality: When the measurement setup itself is
kept constant we can compare the measurements of different constructs
(e.g., different FRET labeling positions or different attachment sites
in a tweezer experiment). The best value $\hat{\alpha}_{o}$ corresponds
to the provably best construct.

Finally, $\hat{\alpha}_{o}$ can be determined by fitting the autocorrelation
function of $\hat{\psi}_{2}$, as described in the spectral estimation
procedure described below. Figs.~\ref{fig:Rate-Estimate}-\ref{fig:Results-Myoglobin}
show estimates of the OQ 
of different observed dynamics (via spectral estimation) and of the estimation quality using other rate estimators.

\subsection*{Markov (state) models (MSMs)}

MSMs have recently gained popularity in the modeling of stochastic
dynamics from molecular simulations \cite{NoeSchuetteReichWeikl_PNAS09_TPT,PrinzEtAl_JCP10_MSM1,VoelzPande_JACS10_NTL9,SwopePiteraSuits_JPCB108_6571,ChoderaEtAl_JCP07}.
MSMs can be understood as a way of implicitly performing rate estimates
via discretizing state space into small substates. Let us consider
a MSM obtained by finely discretizing the observed space $y$ into
bins and estimating a transition matrix $\mathbf{T}(\tau)$ amongst
these bins. We have seen that this procedure approximately solves
the optimization problem of Eq.~(\ref{eq_alpha-optimization}) and
the leading eigenvector of $\mathbf{T}(\tau)$ approximates the projection
of the true second eigenfunction, $\hat{\psi}_{2}^{o}(o)$, available
for the given observable $o$. Ref.~\cite{SwopePiteraSuits_JPCB108_6571}
has suggested to use the implied timescale $\hat{t}_{2}=-\tau/\ln(\hat{\lambda}_{2}(\tau))$
as an estimate for the system's slowest relaxation timescale, and
at the same time for a test which choice of $\tau$ leads to a MSM
with a small approximation error. These implied timescales correspond
to the inverse relaxation rates and therefore the MSM rate estimate
is described by Eq.~(\ref{eq_rate-estimate}) with the choice $\tilde{\psi}_{2}=\hat{\psi}_{2}$.
A sufficiently fine MSM thus serves as an optimal single-$\tau$ rate
estimator as its estimation quality approaches the true OQ $\hat{\alpha}_{o}$ for the observed
signal that is being discretized. However, when this signal has a
poor OQ $\hat{\alpha}_{o}$ since it is poorly separating the slowly-converting
states, there is a substantial rate estimator error according to Eq.~(\ref{eq_rate-error})
that decays slowly with $\tau^{-1}$. This  likely  explains the slow convergence
of implied timescales shown in recent MSM  simulation  studies \cite{SwopePiteraSuits_JPCB108_6571,ChoderaEtAl_JCP07,NoeHorenkeSchutteSmith_JCP07_Metastability,PrinzEtAl_JCP10_MSM1,Bowman_JCP09_Villin}.

\subsection*{Estimating state densities and microscopic transition rates}

When the rate $\kappa_{2}$ is exactly known, the microscopic transition
rates between the two interchanging states, $k_{AB}$ and $k_{BA}$
could be calculated from the equations: 
\begin{eqnarray}
\pi_{A}k_{AB} & = & \pi_{B}k_{BA},\label{eq_detailed-balance}\\
k_{AB}+k_{BA} & = & \kappa_{2}\label{eq_rate-sum}
\end{eqnarray}
where $\pi_{A}$ and $\pi_{B}$ are the stationary probabilities of
states $A$ and $B$:
\begin{eqnarray}
\pi_{A} & = & \int_{o}do\:\mu_{A}^{o}(o)\label{eq_piA-exact}\\
\pi_{B} & = & \int_{o}do\:\mu_{B}^{o}(o)=1-\pi_{B}\label{eq_piB-exact}
\end{eqnarray}
with $\mu_{A}^{o}(o)$ and $\mu_{B}^{o}(o)$ being the partial densities
of states $A$ and $B$ in the observable $o$, respectively.

Here we attempt to estimate both the partial densities, $\mu_{A}^{o}(o)$
and $\mu_{B}^{o}(o)$, and from these the microscopic transition rates
via Eqs.~(\ref{eq_detailed-balance}) and (\ref{eq_rate-sum}). The
difficulty is that $\mu_{A}^{o}(o)$ and $\mu_{B}^{o}(o)$ can significantly
overlap in $o$, both due to the way the order parameter used projects
the molecular configurations onto the observable, and due to noise
broadening of the measurement device. This reveals a fundamental weakness
of dividing-surface approaches. Although a dividing surface estimator
can estimate the rate $\kappa_{2}$ for sufficiently large $\tau$
without bias via Eq.~(\ref{eq_multi-tau-error}), it cannot distinguish
between substates on one side of the barrier, and thus assumes the
partial densities $\mu_{A}^{o}(o)$ and $\mu_{B}^{o}(o)$ to be given
by cutting the full density $\mu^{o}(o)$ at the dividing surface.
When the true partial densities overlap, this estimate can be far
off (compare the curves in \refFigEx{2} panel 5, and \refFigEx{3} panel 5). 
Consequently, incorrect estimates for the microscopic rates $k_{AB}$
and $k_{BA}$ are obtained when Eqs.~(\ref{eq_detailed-balance})
and (\ref{eq_rate-sum}) are used with $\pi_{A}$ and $\pi_{B}$ that
are simply the total densities ``left'' and ``right'' of the dividing
surface.

Hidden Markov models approach this problem by proposing a specific
functional form of $\mu_{A}^{o}(o)$ and $\mu_{B}^{o}(o)$, for example
a Gaussian distribution, and then estimating the parameters of this
distribution with an optimization algorithm \cite{Rabiner_IEEE89_HMM,ChoderaEtAl_BiophysJ11_BHMM}.
This is approach is very powerful when the true functional form of
the partial densities is known, but will give biased estimates when
the wrong functional form is used. 

Here, we propose a nonparametric solution that can estimate the form
of the partial densities $\mu_{A}^{o}$ and $\mu_{B}^{o}$, and the
microscopic transition rates $\hat{k}_{AB}$ and $\hat{k}_{BA}$ in
most cases without bias. For this, we employ the theory of Perron cluster cluster analysis (PCCA+) \cite{DeuflhardWeber_PCCA,KubeWeber_JCP07_CoarseGraining}
which is based on PCCA theory \cite{SchuetteFischerHuisingaDeuflhard_JCompPhys151_146,Deuflhard_LinAlgAppl_PCCA}
which allows for a way of splitting state space into substates and
at the same time maintain optimal approximations to the exact eigenfunctions
(here $\psi_{2}$): The state assignment must be fuzzy, i.e., instead
of choosing a dividing surface that uniquely assigns points $o$ to
either $A$ or $B$, we have fuzzy membership functions $\chi_{A}(o)$
and $\chi_{B}(o)$ with the property $\chi_{A}(o)+\chi_{B}(o)=1$.
These membership functions can be calculated after $\psi_{2}$ is
known. 

In order to compute the membership $\chi_{A}$ and $\chi_{B}$, the
memberships of two points of the observable $o$ must be fixed. The
simplest choice is to propose two observable values that are pure, i.e., that
have a membership of 1 to $A$ and $B$ each. Such an approach is
also proposed by the signal-pair correlation analysis approach \cite{HoffmannWoodside_Angewandte11_PairCorrelationAnalysis}
where the pure values need to be defined by the user. However at this
point of our analysis, an optimal choice can be made, because the
eigenfunction $\hat{\psi}_{2}^{o}$ has been approximated.
Thus we propose to follow the approach of \cite{DeuflhardWeber_PCCA} and
choose the $o$-values where $\hat{\psi}_{2}^{o}$ achieves a minimum
and a maximum, respectively, as purely belonging to $A$ and $B$.
Typically, these are the states the are on the left and the right
boundary of the histogram in $o$. This approach will only start to
give a biased estimate when the overlap of the $A$ and $B$ densities
is so large that not even these extreme points are pure (see \refFigEx{3},
last row for such an example).

{
Let $\hat{\boldsymbol{\psi}}_{2}$ be the second eigenvector of the
Markov model $\mathbf{T}(\tau)$ of the finely binned observable (Eq.
(\ref{eq_1d-transition-matrix})). Then, $\hat{\boldsymbol{\psi}}_{2}$
is a discrete approximation to the projected eigenfunction $\hat{\psi}_{2}^{o}$
. Following the derivation given in the SI, the fuzzy membership
functions on the discretized observable space are given by
\begin{eqnarray}
	\hat{\chi}_{A,i} & = & 
	\frac{\max_{{j}}\hat{{\psi}}_{2,j}-\hat{\psi}_{2,i}}{\max_{{j}}\hat{{\psi}}_{2,j}-\min_{{j}}\hat{{\psi}}_{2,j}}\label{eq_chiA}\\
	\hat{\chi}_{B,i} & = & 
	\frac{\hat{{\psi}}_{2,i}-\min_{{j}}\hat{{\psi}}_{2,j}}{\max_{{j}}\hat{{\psi}}_{2}-\min_{{j}}\hat{{\psi}}_{2}}.\label{eq_chiB}
\end{eqnarray}

where the subscripts $i$ and $j$ denote the discrete state index. 
Note that the extreme values $\max_{{j}}\hat{{\psi}}_{2,j}$ and $\min_{{j}}\hat{{\psi}}_{2,j}$ may have large statistical uncertainties,
when a fine and regular binning is used to discretize the observation. In order to avoid our estimates to be dominated by
statistical fluctuations, we choose the outer most discretization bins such at at least least 0.05\% of the total collected data is in
each of them. The exact choice of this value appears to be irrelevant---as shown in the SI, any choice between 0.005\% and 5\%
of the data yields similar results.}
Since we are restricted to the projected eigenfunction $\hat{\psi}_{2}$,
we can determine the optimal choice $\hat{\chi}_{A}(o)$ and $\hat{\chi}_{B}(o)$
from $\hat{\psi}_{2}(o)$. 

Together with the estimated
stationary density $\mu^{o}(o)$ which can, e.g., be obtained by computing
a histogram from sufficiently long equilibrium trajectories, the probability
of being in $A$ and $B$ is thus given by:
\begin{eqnarray}
\pi_{A} & = & \sum_{i}\mu_{i}^{o}\hat{\chi}_{A,i}\label{eq_piA}\\
\pi_{B} & = & \sum_{i}\mu_{i}^{o}\hat{\chi}_{B,i}=1-\pi_{A}\label{eq_piB} .
\end{eqnarray}

These probabilities can be used to split $\hat{\kappa}_{2}$
into microscopic transition rates $k_{AB}$ and $k_{BA}$:
\begin{eqnarray}
\hat{k}_{AB} & = & \pi_{B}\hat{\kappa}_{2}\label{eq_kAB}\\
\hat{k}_{BA} & = & \pi_{A}\hat{\kappa}_{2}\label{eq_kBA}
\end{eqnarray}
Note that the assignment of labels ``$A$'' and ``$B$'' to parts
of state space is arbitrary. Eq.~(\ref{eq_kAB}) is the transition
rate from $A$ to $B$ as defined by Eqs.~(\ref{eq_chiA})-(\ref{eq_chiB}),
and Eq.~(\ref{eq_kBA}) is the corresponding transition rate from
B to A.

\subsection*{Spectral estimation procedure}

The optimal estimator for $\kappa_{2}$ is thus one that fits the
exponential decay of $\hat{\lambda}_{2}(\tau)$ while minimizing the
fitting error Eq.~(\ref{eq_multi-tau-error}). As analyzed above,
the systematic fitting error is minimized by any multi-$\tau$ estimator.
In order to obtain a numerically robust fit, especially in the case
when statistical noise is present, it is optimal to fit to an autocorrelation
function $\tilde{\lambda}_{2}(\tau)$ where the relevant slowest decay
has maximum amplitude $\hat{\alpha}_{0}$. This is approximately achieved
by constructing a fine discretization MSM on the observed coordinate
(see Section ``Optimal choice of $\tilde{\psi}_{2}$''). Thus, the
optimal estimator of $\kappa_{2}$ proceeds as outlined in (1-4) below.
The full spectral estimation algorithm (1-6) additionally provides
estimates for the microscopic rates $k_{AB}$, $k_{BA}$, and for
the partial densities $\mu_{A}$ and $\mu_{B}$:
\begin{enumerate}
\item { Obtain a fine discretization of the observed coordinate $o$ into
$n$ bins, say $[o_{i},o_{i+1}]$ for $i\in1...n$. When using an equidistant binning make sure to increase
the outer most states to a size to cover a significant part (e.g. 0.05\%) of the total population.}
\item Construct a row-stochastic transition matrix $\mathbf{T}(\tau)$ for
different values of $\tau$. The estimation of transition matrices
from data have been described in detail \cite{PrinzEtAl_JCP10_MSM1}.
A simple way of estimating $\mathbf{T}(\tau)$ is the following: (i)
for all pairs $i,j$ of bins, let $c_{ij}(\tau)$ be the number of
times the trajectory has been in bin $i$ at time $t$ and in bin
$j$ at time $t+\tau$, summed over all time origins $t$; (ii) estimate
the elements of $\mathbf{T}(\tau)$ by $T_{ij}(\tau)=c_{ij}(\tau)/\sum_{k}c_{ik}(\tau)$.
A numerically superior approach is to use a reversible transition
matrix estimator \cite{PrinzEtAl_JCP10_MSM1}.
\item Calculate the discrete stationary probability $\boldsymbol{\mu}$
and the discrete eigenvector $\hat{\boldsymbol{\psi}}_{2}$ by solving
the eigenvalue equations:
\begin{eqnarray}
\mathbf{T}^{T}(\tau)\boldsymbol{\mu} & = & \boldsymbol{\mu}\label{eq}\\
\mathbf{T}(\tau)\hat{\boldsymbol{\psi}}_{2} & = & \hat{\lambda}_{2}\hat{\boldsymbol{\psi}}_{2}\label{eq:-2}
\end{eqnarray}
with the largest eigenvalues $\lambda_{1}=1<\hat{\lambda}_{2}\le\hat{\lambda}_{3}$.
$\mathbf{T}^{T}$ denotes the transpose of the transition matrix.
The $i$-th element of the vectors $\boldsymbol{\mu}$ and $\hat{\boldsymbol{\psi}}_{2}$
approximate the stationary density $\mu(o)$ and $\hat{\psi}_{2}$
on the respective point $o=\frac{o_{i}+o_{i+1}}{2}$. Functions $\mu^{o}(o)$
and $\psi^{o}_{2}(o)$ can be obtained by some interpolation method.
\item Estimate the relaxation rate $\hat{\kappa}_{2}$ and the OQ $\hat{\alpha}$
via an exponential fit of $\alpha e^{-\kappa_{2}\tau}$ to the tail
of $\hat{\lambda}_{2}(\tau)=\langle\hat{\psi}_{2}(t)\hat{\psi}_{2}(t+\tau)\rangle_{t}$.

\item Calculate the partial densities $\boldsymbol{\mu}_{A}$ and $\boldsymbol{\mu}_{B}$
from Eqs.~(\ref{eq_piA}) and (\ref{eq_piB}) using transition matrix
eigenvectors estimated at a lag time $\tau_{min}$ at which the rate
estimate $\hat{\kappa}_{2}$ is converged.

\item Calculate the microscopic transition rates $k_{AB}$ and $k_{BA}$
from Eqs.~(\ref{eq_kAB}) and (\ref{eq_kBA})
\end{enumerate}
Note that this estimator is optimal in terms of minimizing the systematic
error. When dealing with real data, the amount of statistics may set
restrictions of how fine a discretization is suitable and how large
a lag time $\tau$ will yield reasonable signal to noise. For a discussion
of this issue refer to e.g., Ref.~\cite{ZwanzigAilawadi_PRE69_StatisticalErrorCorrelations}.

\subsection*{Illustrative two-state example}

\begin{figure*}[t]
\noindent \begin{centering}
	\includegraphics[width=0.8\textwidth]{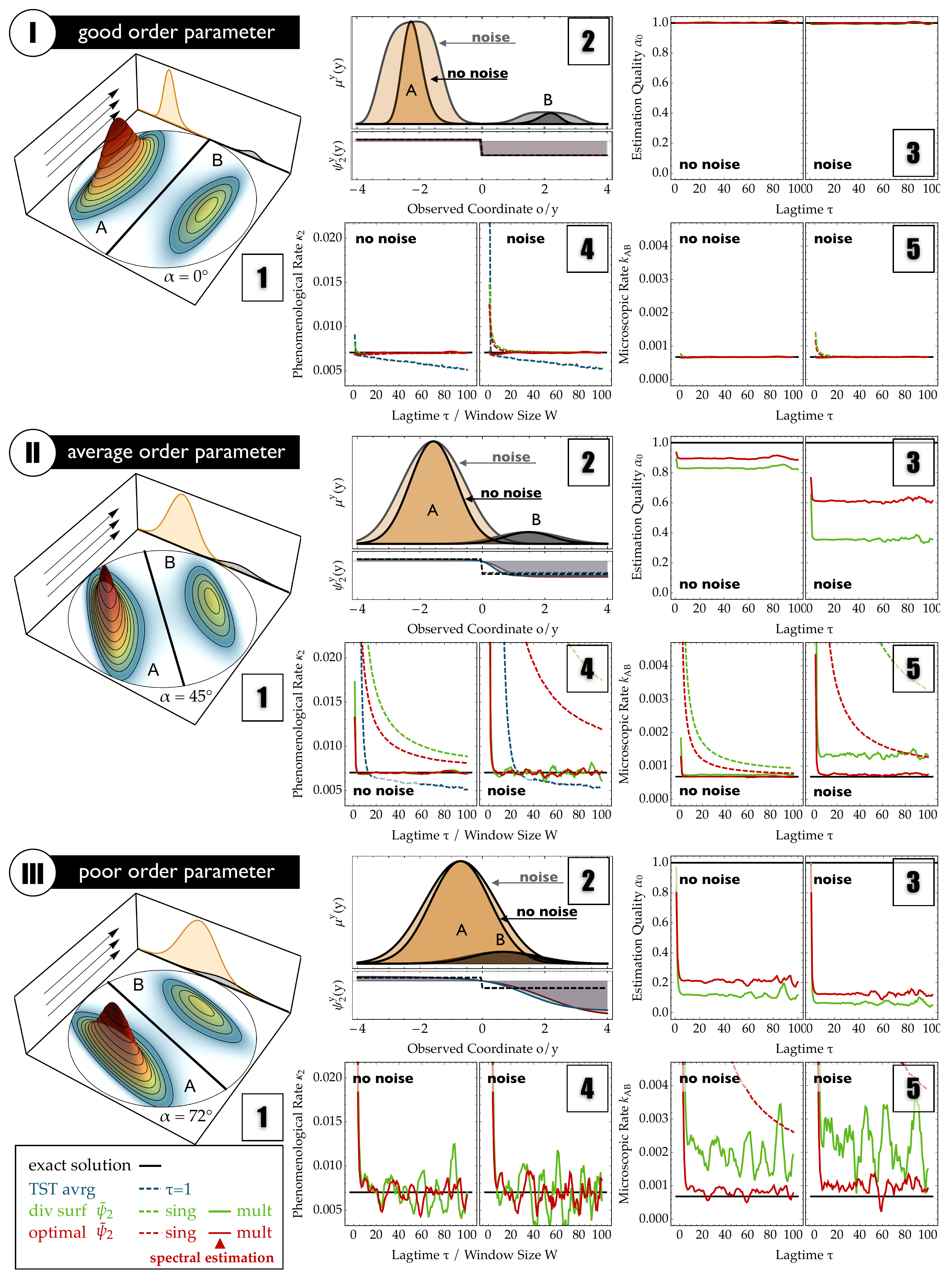}
\par\end{centering}

\caption{\textcolor{blue}{\label{fig:Rate-Estimate} }Estimation results using
overdamped Langevin dynamics in a two-dimensional two-well potential
that is projected onto different observables: (I) perfect projection,
(II) average-quality projection, (III) poor projection. Results
are compared without noise (left half of panels) and with additional measurement
noise (right half of panels). 
(1) Full state space with indicated direction of the used order parameter.
{ (2) (top) Stationary density $\mu^{y}(y)$ in the observable of the two partial densities of states $A$ (orange) and $B$ (gray). Results with noise are shaded lighter and are more spread out. (bottom) Second eigenvector without noise (solid, blue), with noise (solid, red) and dividing surface (black, dashed).}
(3) Estimation quality $\alpha$ from spectral estimation (OQ, red), and
from exponential fitting to the number correlation function using
a diving surface at $y=0$ (green). 
(4) Estimated relaxation rate $\kappa_{2}$: TST with averaging window of size W (indicated in the x-axis).
Dividing surface at $o=0$ with single-$\tau$ (dashed green line) and multi-$\tau$ (solid green line) estimators.
Estimates from an MSM-derived second eigenvector $\tilde{\psi}_2$ with a single-$\tau$ estimate (normal MSM, dashed red line),
and multi-$\tau$ estimate (spectral estimation, solid red line). 
The black line is the reference solution, obtained from a direct MSM estimate for $\tau=50$ in row 1. 
(5) The transition rates $k_{AB}$ from state $A$ to $B$. The coloring is identical to panels (4).}
\end{figure*}

To illustrate the theory and the concepts of this paper, we compare
the behavior of different order parameters, measurement noise and
different estimators in Fig.~\ref{fig:Rate-Estimate}. The full-space
model here is a two-dimensional model system using overdamped Langevin
dynamics in a bistable potential. This choice was made because the
exact properties of this system are known and the quality of different
estimates can thus be assessed. The potential is chosen such that
the eigenfunction associated with the slow process, $\psi_{2}(\mathbf{x})$
varies in $x_{1}$ and is constant in $x_{2}$, such that the choice
$o=x_{1}$ represents a perfect projection and the choice $o=x_{2}$
represents the worst situation in which the slow process is invisible.

{
Fig.~\ref{fig:Rate-Estimate} shows three scenarios using

\renewcommand{\theenumi}{\Roman{enumi}}
\begin{enumerate}
\item $y=x_{1}$ \newline (perfect order parameter --- projection angle $0\text{\textdegree}$),
\item $y=\frac{1}{2}(x_{1}+x_{2})$ \newline (average order parameter - projection angle $45\text{\textdegree}$), and 
\item $y=\frac{1}{4}(x_{1}+3x_{2})$ \newline (poor order parameter --- projection angle $72\text{\textdegree}$).
\end{enumerate} Additionally, we compare the results when the order parameter $y$
is traced without noise (left half of panels 3-5), and when measurement noise
is added (right half of panels 3-5). Noise here consists of adding a uniformly
distributed random number from the interval {[}-1,1{]} to the signal,
such that the noise amplitude is roughly 25\% of the signal amplitude.
}

Fig.~\ref{fig:Rate-Estimate}, panels 2 show the apparent
stationary density in the observable $y$, $\mu^{y}(y)$, or in the
noisy observable $o$, $\mu^{o}(o)$, as a black solid line. The partial
densities of substates $A$ (orange) and $B$ (grey) which comprise
the total stationary density are shown as well. The lower part of
the figure shows the observed eigenfunction associated with the two-state
transition process ($\psi_{2}^{y}$ or $\psi_{2}^{o}$) as black solid
line with grey background. { For comparison the results in the case of noise are
shown in the background with lighter colors.} It is apparent that when the quality of
the observation is reduced, either by choosing a poor order parameter,
or by adding experimental noise, the overlap of the partial densities
increases and the continuous projected eigenfunction $\psi_{2}^{o}$
becomes smoother and thus increasingly deviates from the dividing
surface model which is a step function switching at the dividing surface
(dashed line).

Panels 3 show the estimation qualities or observation qualities (OQ) in these
different scenarios. The fact that the green and red lines are approximately
constant after $\tau=5$ (when the fast processes have relaxed) shows
that the OQ can be reliably estimated at these lag time ranges using
either the dividing surface or the spectral estimation approach. The
red line (spectral estimation) corresponds to the OQ,
which varies between 1 (perfect order parameter I) and 0.15 (poor
order parameter III with additional measurement noise). It is seen
that the OQ given by the spectral estimator can be much larger
than the suboptimal estimation quality of the dividing surface estimator
that uses a fit to the number correlation function Eq. (\ref{eq_number-correlation-function-1-1})
(green line). This is especially apparent in the case of an intermediate-quality
order parameter (Fig.~\ref{fig:Rate-Estimate}II-3).

{
Panels 4 show the estimate of the relaxation rate $\kappa_{2}$
obtained for the three scenarios where each panel compares 5 different rate estimators with the exact result (black solid line):
(1) Direct counting of transitions from time-filtered data (TST estimate, blue line). For this estimator, the x-axis denotes the length of the averaging window $W$, 
ranging from 1 to 100 frames. 
(2,3) The dividing surface estimates using either a single-$\tau$ estimator (\ref{eq_rate-estimate}) (dashed green line), and the multi-$\tau$ estimator (solid green line).
(4,5) The single-$\tau$ MSM estimate (dashed red line), and the multi-$\tau$ MSM estimate (spectral estimation, solid red line).
For the single-$\tau$ and the exact estimators the x-axis indicates the used lag time $\tau$ in the estimation where for the multi-$\tau$ estimators (i.e., dividing surface and spectral estimation), the x-axis specifies $\tau$ which is the start of the time range $[\tau$, $\tau+10]$ used for an exponential fit. 

In the case of a perfect order parameter (I), all estimators
yield the correct rate at lag times $\tau>5$ time steps (where the
fast processes with rates $\kappa_{3}$ or greater have disappeared).
Only in the case of TST (blue line) with increasing size $W$ of the filtering window, the estimated
rate tends to be too slow because an increased number of short forward-and-backward transition events
become smeared out by the filtering window, therefore systematically underestimating the rate.
For the perfect order parameter I, the noise has little effect on the estimate, because
the partial densities of states $A$ and $B$ are still well separated.

For the average-quality and poor order parameters, the MSM estimate
breaks down dramatically, providing a strongly overestimated rate
for $0<\tau<100$ time steps. Panels II-4 and III-4 show the
typical behavior of the $\tau^{-1}$-convergence of the MSM estimate predicted
by the theory (Eq.~(\ref{eq_rate-error})). Clearly, the MSM estimate
will converge to the true value for very large values of $\tau$,
but especially for the situation of a poor order parameter, the minimal
$\tau$ required to obtain a small estimation error is larger
than the timescale $\kappa_{2}^{-1}$ of the slowest process, thus
rendering a reliable estimation impossible. 

It is seen that the magnitude of the error for a given value of $\tau$ increases when either adding noise 
(left half of panels 4 vs right half) or decreasing the quality of the order parameter
(panel II-4 vs panel III-4). This is because in this sequence the OQ deteriorates, as predicted
by the theory of reaction coordinate qualities (see above), and hence
the prefactor of the MSM error increases (see Eq.~(\ref{eq_rate-error})).

As predicted by Eq.~(\ref{eq_mult-is-better-than-single}), the multi-$\tau$
estimators (dividing surface and spectral estimate, red and green solid lines) are always better
than the single-$\tau$ estimates (red and green dashed lines). 
As predicted by Eq.~(\ref{eq_multi-tau-error}),
both the dividing surface and spectral estimate of $\hat{\kappa}_{2}$
converge when the fast processes have died out (here approximately
at $\tau>5$ time steps). Also, panels II-4 and III-4 show that
the spectral estimate is more stable than the dividing surface estimate,
i.e., it exhibits weaker fluctuations around the true value $\kappa_{2}$.
This is because the spectral estimate uses the OQ $\hat{\alpha}_{o}$ as estimation quality, which is larger than the estimation quality of other estimators,
and thus the exponential tail of the autocorrelation function can
be fitted using a larger amplitude of the process relaxation with
rate $\kappa_{2}$, achieving a better signal-to-noise ratio.

Panels 5 show the results of the microscopic rate $k_{AB}$ 
that quantifies the rate at which rare transition events between the 
large (orange) state $A$ and the smaller (gray) state $B$. The solid lines
indicate the estimates from Eq.~(\ref{eq_kAB}) either using the partial 
densities from the dividing surface (green), or PCCA+ (spectral estimate, red).
For corresponding rates computed from a MSM using the different projections
are shown in dashed lines. As expected, the partial densities from the dividing
surface estimate are significantly biased as soon as the states overlap in the observable,
either due to choosing a poor order parameter, or to experimental
noise. As a result, the dividing surface estimates for the microscopic
rates $k_{AB}$ and $k_{BA}$ are biased for all these cases (panels
II-5, panels III-5). The spectral estimate gives an unbiased
estimate for average overlap (panels II-5). For strong overlap,
even the spectral estimator has a small bias because no pair
of observable states can be found that are uniquely assignable to
states $A$ and $B$. 
Still, the spectral estimator yields good results even in the poor order parameter setting 
(panel III-5). Like for the relaxation
rate ($\kappa_{2}$) estimate, the spectral estimator exhibits less fluctuations here because the larger 
estimation quality yields a better signal-to-noise ratio.
}

\subsection*{Applications to optical tweezer data}

\begin{figure*}[t]
\noindent 
\begin{centering}
\includegraphics[width=1\textwidth]{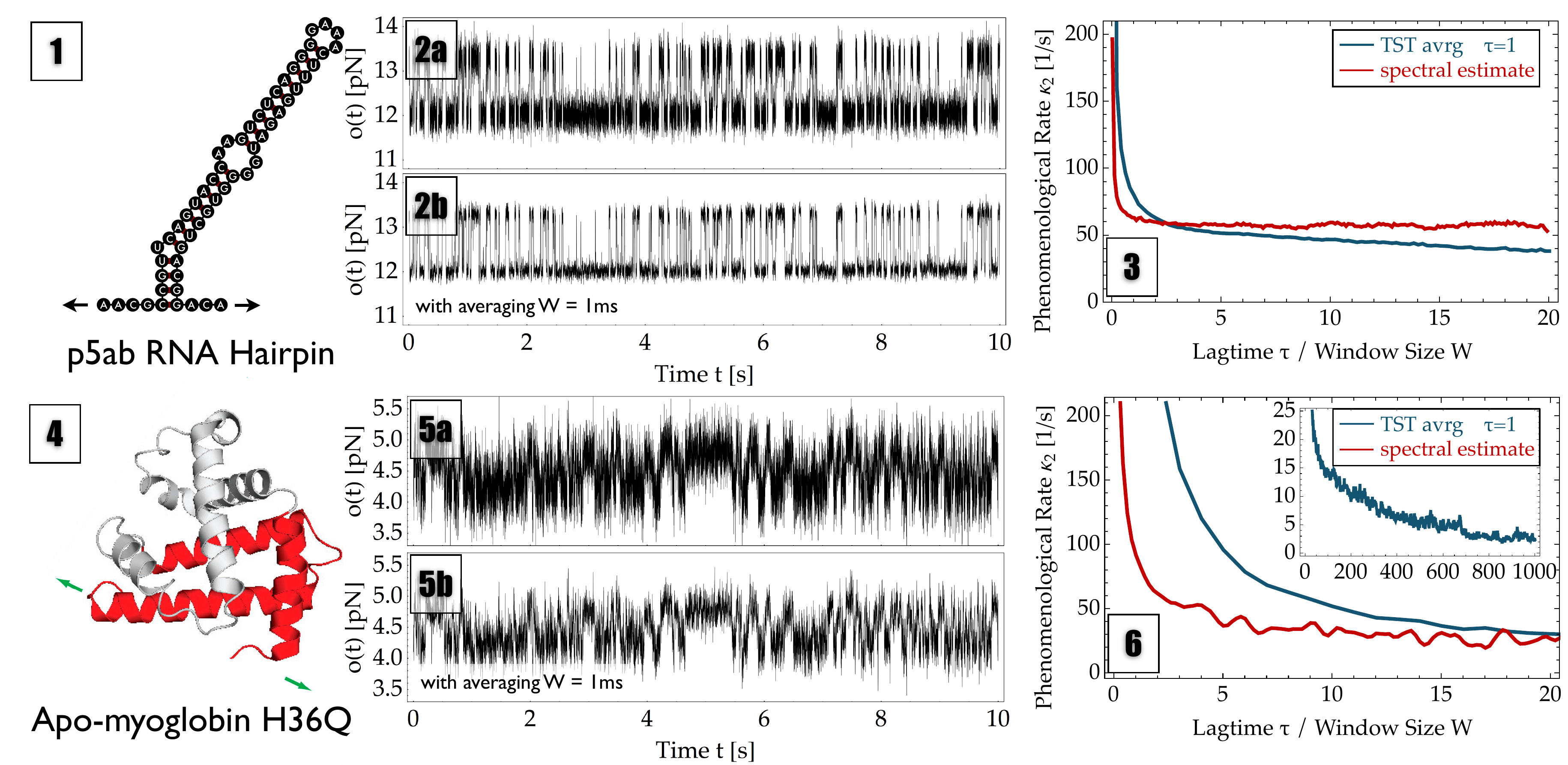}
\par
\end{centering}
\caption{{\label{fig:Experiments_Intro} } {Probed systems by optical tweezer experiments: (top) RNA hairpin p5ab, (bottom) H36Q mutant of sperm whale apo-myoglobin. Panels (1,4) show schematic views of the probed system in its native secondary/tertiary structure including the direction of the pulling force (green and black arrows). Panels (2,5) show the traces used for analysis. Row (a) reports the results when directly analyzing the measured 50 kHz data, while row (b) reports the results when analyzing data that has been binned to 1 kHz to reduce the noise amplitude. Panel (3,6) show the estimated phenomenological rates $\hat{\kappa}_{2}$ for TPT (blue) using different averaging window sizes $W$ (x-axis) and spectral estimation (red) for different lag times $\tau$ (x-axis). For apo-myoglobin the inset displays the long-behavior of TST at large window sizes $W$, where the rate is systematically underestimated.
}}
\end{figure*}

In order to illustrate the performance of spectral estimation on real
data, it is applied to optical tweezer measurements of the extension
fluctuations of two biomolecules examined in a recent optical force
spectroscopy study: the p5ab RNA hairpin \cite{elms:2012:biophys-j:force-feedback}
and the H36Q mutant of sperm whale apo-myoglobin at low pH \cite{elms:2012:pnas:apo-myoglobin}.
The p5ab hairpin forms a stem-loop structure with a bulge under native
conditions (Fig.~\ref{fig:Experiments_Intro}-1) and zips/unzips repeatedly
under the conditions used to collect data (Fig.~\ref{fig:Experiments_Intro}-2a),
while apo-myoglobin (crystal structure shown in Fig.~\ref{fig:Experiments_Intro}-4)
hops between unfolded and molten globule states at the experimental
pH of 5 (Fig.~\ref{fig:Experiments_Intro}-5a) \cite{elms:2012:pnas:apo-myoglobin}.

Experimental force trajectory data were generously provided by the
authors of Refs.~\cite{elms:2012:biophys-j:force-feedback,elms:2012:pnas:apo-myoglobin}.
Experimental details are given therein,
but we briefly summarize aspects of the apparatus and experimental
data collection procedure relevant to our analysis.

The instrument used to collect both datasets was a dual-beam counter-propagating
optical trap \cite{bustamante-smith:2006:minitweezers-patent}. The
molecule of interest was tethered to polystyrene beads by means of
dsDNA handles, with one bead suctioned onto a pipette and the other
held in the optical trap. A piezoactuator controlled the position
of the trap and allowed position resolution to within 0.5 nm, with
the instrument operated in passive (equilibrium) mode such that the
trap was stationary relative to the pipette during data collection.
The force on the bead held in the optical trap was recorded at 50
kHz, with each recorded force trajectory 60 s in duration.

It is common practice to estimate rates in such data by directly counting the number of transitions across some user-defined dividing surface, and dividing
by the total trajectory length. Often, this procedure is applied after filtering the data with a time-running average. 
The results of this common procedure (effectively a TST estimate, or an MSM estimate with $\tau=1$), is shown in
Figs.~\ref{fig:Experiments_Intro}-3,6 (blue line) using various averaging window sizes $W$, and compared to the optimal estimator (spectral estimation)
for a range of estimation lag times $\tau$. Although the TST estimate shows less fluctuations the spectral
estimation result converges much faster and provides a more stable result in terms of the varying parameter (lag time $\tau$ / window size $W$). TST also tends to underestimate the true rate for large window sizes $W$. Moreover, the TST estimate never shows any plateau, thereby making it impossible to decide which rate
estimate should be used.

\subsection*{p5ab RNA Hairpin analysis}
\begin{figure*}[t]
\noindent \begin{centering}
\includegraphics[width=1\textwidth]{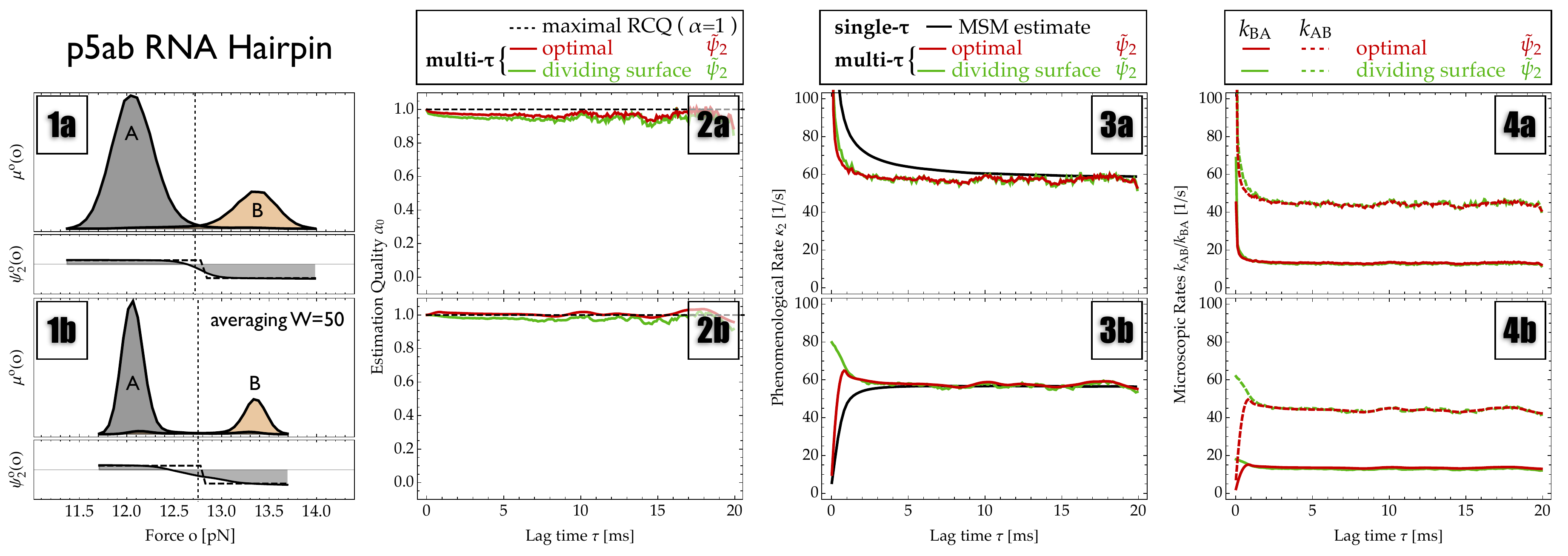}
\par\end{centering}

\caption{\label{fig:Results-p5ab}{Estimates for rates and estimation qualities
from passive-mode single-molecule force probe experiments
of the p5ab RNA hairpin. All panels report the estimation results,
showing the direct MSM estimate (black), a fit to the fluctuation
autocorrelation function using a dividing surface at $o=12.80\:pN$ 
(green), and spectral estimation (red). (1a+b): the stationary probability of observing a given force value
(solid black line). The partial probabilities of states A (grey) and
B (orange) obtained by spectral estimation show that there is very
little overlap between the states. (2a+b): the estimation quality $\alpha_{o}$, 
coinciding with the observation quality (OQ) $\hat{\alpha_{o}}$ for spectral estimation.
(3a+b): estimated relaxation
rate $\kappa_{2}$. (4a+b): estimated microscopic transition rates:
the folding rate $k_{AB}$ (dashed) and the unfolding rate $k_{BA}$
(solid). }}
\end{figure*}

Fig.~\ref{fig:Results-p5ab} compares the results of several  
rate estimators for optical tweezer measurement of the p5ab RNA hairpin
extension fluctuations. 
A sketch of the RNA molecule and the experimental trajectory analyzed can be found in Fig.~\ref{fig:Experiments_Intro}-1,2 top. The trajectory exhibits a
two-state like behavior with state lifetimes on the order of tens
of milliseconds. Fig.~\ref{fig:Results-p5ab}-1a shows the stationary
probability density of measured pulling forces, exhibiting two nearly
separated peaks. Fig.~\ref{fig:Results-p5ab}-2a shows the estimation quality $\alpha_{o}$
(OQ $\hat{\alpha}_{o}$ for the spectral estimator), which is approximately
constant at lag times $\tau>5\:ms$, indicating a reliable estimate
for this quantity at lag times greater than $5\:ms$. An optimum value
of $\hat{\alpha}_{o}\approx0.96$ (spectral estimator) is found while
the best possible dividing surface results in $\alpha_{o}\approx0.94$.
These values indicate that the present reaction coordinate is well
suited to separate the slowly interconverting states, and that different
approaches, including a Markov model, a dividing surface estimate
and a spectral estimate should yield good results.

Fig.~\ref{fig:Results-p5ab}-3a compares the estimates of the relaxation
rate $\kappa_{2}$ using the direct MSM estimate (black), a fit to
the fluctuation autocorrelation function using a dividing surface
at the histogram minimum $o=12.80\:pN$ (green), and spectral estimation
(red). For the multi-$\tau$ estimators (dividing surface and spectral
estimation), the lag time $\tau$ specifies the start of a time range
{[}$\tau$, $\tau+2.5\:ms${]} that was used for an exponential fit.
All estimators agree on a relaxation rate of about $\hat{\kappa}_{2}\approx58\: s^{-1}$,
corresponding to a timescale of about $17\: ms$. The MSM estimate
is strongly biased for short lag times, exhibiting the slow $\tau^{-1}$
convergence predicted by the theory for single-$\tau$ estimators
(Eq.~(\ref{eq_rate-error})). It converges to an estimate within
10\% of the value from multi-$\tau$ estimates after a lag time of
about $10\: ms$. The dividing surface and spectral estimators behave
almost identical, and converge after about $\tau=5\: ms$. According
to the error theory of multi-$\tau$ estimators (Eq.~(\ref{eq_multi-tau-error})),
this indicates that there are additional, faster kinetics in the data,
the slowest of which have timescales of $2-3\:ms$. In agreement with
the theory (Eq.~(\ref{eq_mult-is-better-than-single})), the multi-$\tau$
estimators (dividing surface, spectral estimate) converge faster than
the single-$\tau$ estimate (MSM).

As indicated in Fig.~\ref{fig:Results-p5ab}-1a, the substates estimated
from PCCA+ are almost perfectly separated and can be well distinguished
by a dividing surface at the histogram minimum $o=12.80\:pN$ . Consequently,
both the dividing surface estimate and the spectral estimate yield
almost identical estimates of the microscopic transition rates --- the
folding rate being $k_{AB}\approx45\: s^{-1}$ and the unfolding rate
being $k_{BA}\approx15\: s^{-1}$ (Fig.~\ref{fig:Results-p5ab}-4).
In summary, the two-state kinetics of p5ab can be well estimated by
various different rate estimators because the slowly-converting states
are well separated in the experimental observable.

Fig.~\ref{fig:Results-p5ab} panels b show estimation results for
data that has been filtered by averaging over 50 frames ($1\:ms$). This
averaging further reduces the already small overlap between substates
$A$ and $B$ while the filter length is much below the timescale
of $A-B$ interconversion. Therefore, filtering has a positive result
on the analysis: The effective OQ $\hat{\alpha}_{o}$ increases and
is now approximately equal to 1 according to spectral estimation.
The estimation results are largely identical to the case with noise.
In Fig.~\ref{fig:Results-p5ab}-3b, the error made by the Markov
model estimate has become smaller because the error prefactor reported
in Eq.~(\ref{eq_rate-error}), $\ln\alpha_{o}$, has become smaller.
Note that in contrast to the unfiltered data analysis, some of the
rate estimates (MSM and spectral estimate) underestimate the rate
for small lag times $\tau$. This is not in contradiction with our
theory which predicts an overestimation of the rate for Markovian
processes. By using the filter, one has effectively introduced memory
into the signal, and the present theory will only apply at a lag time
$\tau$ that is a sufficiently large multiple of the filter length,
such that the introduced memory effects have vanished.

\subsection*{Apo-myoglobin analysis}
 
\begin{figure*}[t]
\noindent \begin{centering}
\includegraphics[clip,width=1\textwidth]{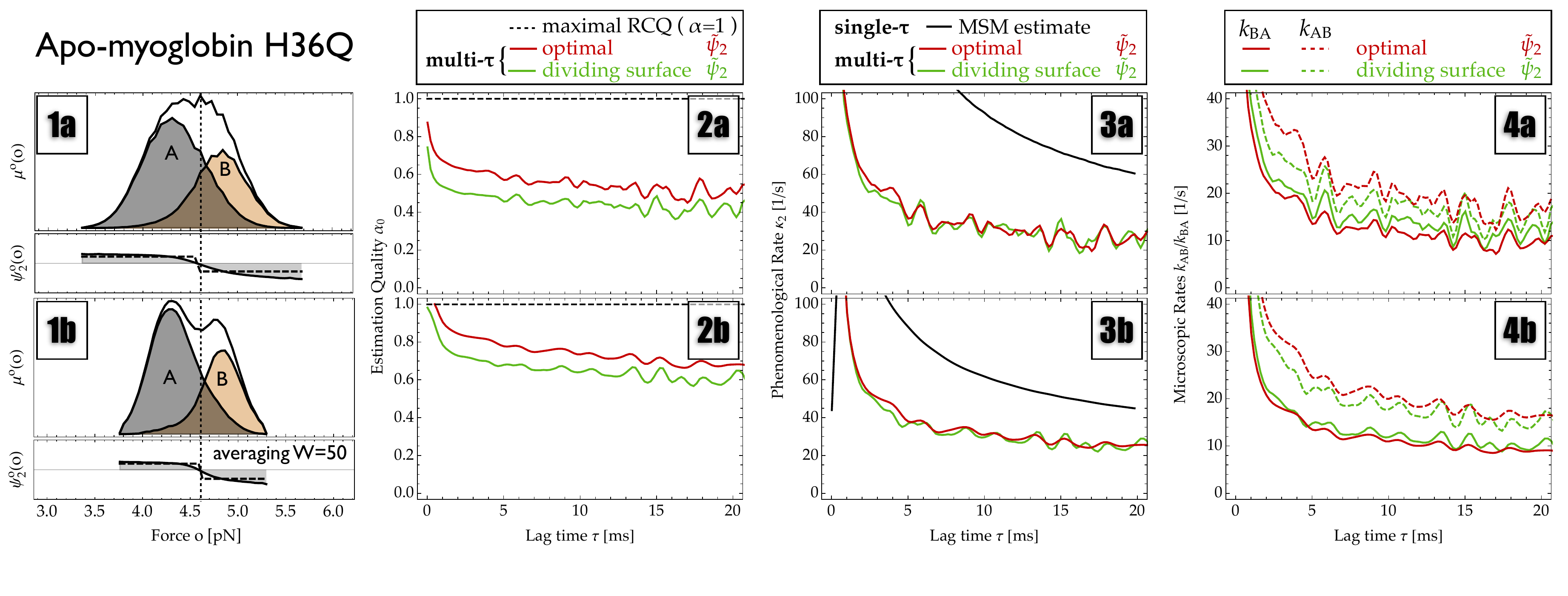}
\par\end{centering}

\caption{{\label{fig:Results-Myoglobin}}{Estimates for rates
and estimation qualities from passive-mode single-molecule
force probe experiments of apo-myoglobin. All panels report the
estimation results, showing the direct Markov model estimate (black),
a fit to the fluctuation autocorrelation function using a dividing
surface at the histogram maximum {(minimum between two maxima for filtering)} $o=4.6\:pN$  (green), and spectral
estimation (red). (1a+b): the stationary probability of observing a given force value
(solid black line). The partial probabilities of states A (grey) and
B (orange) obtained by spectral estimation show that there is very
little overlap between the states. (2a+b): the estimation quality $\alpha_{o}$, 
coinciding with the observation quality (OQ) $\hat{\alpha}_{o}$ for spectral estimation.
(3a+b): estimated relaxation rate $\kappa_{2}$. (4a+b): estimated microscopic transition rates
$k_{AB}$ (dashed) and $k_{BA}$ (solid). }}
\end{figure*}

Fig.~\ref{fig:Results-Myoglobin} shows estimation results for an
optical tweezer experiment that probes the extension fluctuations
of apo-myoglobin \cite{ElmsChoderaBustamanteMarqusee_PNAS12_ApoMyoglobin}.
Fig.~\ref{fig:Experiments_Intro}-4 shows a sketch of the experimental
pulling coordinate (green arrows) depicted at the crystal structure
of apo-myoglobin. Fig.~\ref{fig:Experiments_Intro}-5 shows the trajectory
that was analyzed. Out of the trajectories reported in \cite{ElmsChoderaBustamanteMarqusee_PNAS12_ApoMyoglobin}
we have here chosen one where the two slowest-converting states have
a large overlap. While the trajectory indicates that there are
at least two kinetically separated states, the stationary probability
density of measured pulling forces (Fig.~\ref{fig:Results-Myoglobin}-1a)
does not exhibit a clear separation between these states in the measured
pulling force. This is also indicated by Fig.~\ref{fig:Results-Myoglobin}-2a,
which shows that the optimal observation quality (OQ) has
a value of $\hat{\alpha}_o \approx 0.5$ (spectral estimator) at $\tau=15\:ms$
while the best possible dividing surface results only yield an estimation quality of $\alpha_{o}\approx0.4$
at $\tau=15\:ms$. Thus, the quality of the apo-myoglobin data is similar
to that of the two-state model with intermediate-quality order parameter
and noise (Fig.~\ref{fig:Rate-Estimate}II-b). This data thus represents
a harder test for rate estimators than the p5ab hairpin and should
show differences between different rate estimators.

Fig.~\ref{fig:Results-Myoglobin} panels 3 and 4 compare the estimates
of $\kappa_{2}$ from the direct MSM estimate (black), a fit to the
fluctuation autocorrelation function using a dividing surface at the
local histogram maximum {(minimum between two maxima with filtering)} of the binned data at $o=4.6\:pN$  (green),
and spectral estimation (red). For the multi-$\tau$ estimators (dividing
surface and spectral estimation), the lag time $\tau$ specifies the
start of a time range {[}$\tau$, $\tau+2.5\:ms${]} that was used
for an exponential fit. 

Fig.~\ref{fig:Results-Myoglobin}-3a shows again, that the MSM estimate
of $\kappa_{2}$ exhibits the slow $\tau^{-1}$ convergence predicted
by the theory (Eq.~(\ref{eq_rate-error})) and does not yield a converged
estimate using lag times of up to $20\: ms$. Since the MSM estimate
still significantly overestimates the rate at $\tau=50\: ms$ when
the relaxation process itself has almost entirely decayed, this estimator
is not useful to analyze the apo-myoglobin data. In contrast, both,
the dividing surface multi-$\tau$ approach and the spectral estimator
do yield a converged estimate of $\hat{\kappa}_{2}\approx26\: s^{-1}$,
corresponding to a timescale of about $38\: ms$ (Fig.~\ref{fig:Results-Myoglobin}-3a).
In Ref. \cite{ElmsChoderaBustamanteMarqusee_PNAS12_ApoMyoglobin}
a Hidden Markov model with Gaussian output functions was used and
the rate was estimated to be $\hat{\kappa}_{2}\approx46\: s^{-1}$
corresponding to a timescale of approximately $21\: ms$. These
differences are consistent with our theory which shows that rate estimation
errors lead to a systematic overestimation of the rate (and underestimation
of the timescale). 
Fig.~\ref{fig:Results-Myoglobin}-1a shows the possible reason why the Gaussian HMM
in \cite{ElmsChoderaBustamanteMarqusee_PNAS12_ApoMyoglobin} yields a rate overestimate:
the partial probabilities are clearly not Gaussians.
Following our theory, the smallest rate estimates the best estimates, which
are here provided by the multi-$\tau$ estimators using either dividing
surface or spectral estimation approaches.

In agreement with the theory (Eq.~(\ref{eq_mult-is-better-than-single})),
the multi-$\tau$ estimators (dividing surface, spectral estimate)
converge faster than the single-$\tau$ estimate (MSM). 
A double-exponential fit to the spectral estimation autocorrelation function yields an 
estimate of $\kappa_3 \approx 100 s^{-1}$, corresponding to a timescale of $10\:ms$. 
Thus there is a timescale separation of a factor of about 4 between the slowest and the 
next-slowest process, indicating that when viewed at sufficiently large timescales ($> 20\:ms$), 
the dynamics can be considered to be effectively two-state.
However, since the present of faster processes is clearly visible in the data, 
it may be worthwhile to investigate further substates of the $A$ and
$B$ states with multistate approaches such as Hidden Markov Models
\cite{ChoderaEtAl_BiophysJ11_BHMM} or pair correlation analysis \cite{HoffmannWoodside_Angewandte11_PairCorrelationAnalysis}.
Such an analysis is beyond the scope of the present paper on two-state rate theory.

As indicated in Fig.~\ref{fig:Results-Myoglobin}-1a, the substates
$A$ and $B$ estimated from PCCA+ do strongly overlap. Thus, even
though the dividing surface estimator can recover the true relaxation
rate $\kappa_{2}$, the estimated microscopic rates $k_{AB}$ and
$k_{BA}$ depend on the choice of the position of the dividing
surface. Fig.~\ref{fig:Results-Myoglobin}-4a shows the estimates
the dividing surface multi-$\tau$ estimator, evaluating to $k_{AB}\approx12\: s^{-1}$
and $k_{BA}\approx15\: s^{-1}$. In contrast, the spectral estimator
yields estimates of $k_{AB}\approx 16\: s^{-1}$ and $k_{BA}\approx10\: s^{-1}$. 
Although not being strongly different the diving surface approach suggests 
a reversed dominant direction of the process.

Like for the two-state model results shown in Fig.~\ref{fig:Rate-Estimate},
the spectral estimate is numerically more stable in $\tau$ compared
to the dividing surface estimate as a result of achieving a better
signal-to-noise ratio. Clearly, in the dividing surface approach it
is possible to pick a dividing surface position which yields the same
estimates for $k_{AB}$ and $k_{BA}$ like the spectral estimator.
However, the dividing surface estimator itself does not provide any
information which is the correct choice, and therefore this theoretical
possibility is of no practical use. Supplementary Figure 2 compares
the estimation results of $\kappa_{2}$ and $k_{AB}$, $k_{BA}$ for
different choices of the dividing surface. In contrast to the dividing
surface approach, the spectral estimator only assumes that the extreme
values of $o$ are pure, which is a much weaker requirement than assuming
that an appropriate dividing surface exists (see theory), and hence
provides more reliable rate estimates.

The ``b''-panels in Fig.~\ref{fig:Results-Myoglobin} show the
effect of filtering the data on the estimation results. Here, the
data was averaged over a window length of $1\: ms$, corresponding
to an averaging of 50 data points of the original 50 kHz data. Fig.~\ref{fig:Results-Myoglobin}-1b
indicate that this filtering enhances the separation of states, and
the apparent OQ increases to about $\hat{\alpha}_{0}\approx0.7$
(spectral estimate) while the dividing surface estimation quality is $\alpha_{0}\approx0.6$.
The relaxation rate $\kappa_{2}$ is still estimated to have $\hat{\kappa}_{2}\approx26\: s^{-1}$
and the estimate becomes more robust for both the dividing surface
and spectral estimates (Fig.~\ref{fig:Results-Myoglobin}-3b). The
MSM estimate slightly improves but is still significantly too high.
{ Fig.~\ref{fig:Results-Myoglobin}-4b shows that the dividing surface
derived rate estimates $k_{AB}$ and $k_{BA}$ have improved and are now
similar to the spectral estimation results, while
the spectral estimate itself remains at $k_{AB}\approx 16\: s^{-1}$ and $k_{BA}\approx10\: s^{-1}$
independent of the filtering, which is in support of the reliability
of the spectral estimate.}

\subsection*{Summary}

We have described a rate theory for observed two-state dynamical systems.
The underlying system is assumed to be ergodic, reversible, and Markovian
in full phase space, as fulfilled by most physical systems in thermal
equilibrium. The observation process takes into account that the system
is not fully observed, but rather by tracing one order parameter (the
extension to multiple oder multidimensional order parameters is straightforward).
During the observation process, the observed order parameter may be
additionally distorted or dispersed, for example by experimental noise.
Such observed dynamical systems occur frequently in the molecular
sciences, and appear both in the analysis of molecular simulations
as well as of single-molecule experiments.

The presented rate theory for observed two-state dynamics is a generalization
to classical two-state rate theories in two ways: First, most available
rate theories assume that the system of interest is either fully observable,
or the relevant indicators of the slowest kinetic process can be observed
without projection error or noise. Secondly, most classical rate theories
are built on specific dynamical models such as Langevin or Smoluchowski
dynamics. The present theory explicitly allows the two kinetic states
to overlap in the observed signal (either due to using a poor order
parameter or to noise broadening), and does not require a specific
dynamical model, but rather works purely based on the spectral properties
of a reversible ergodic Markov propagator --- hence the name spectral
rate theory.

Given the spectral rate theory, the systematic errors of available
rate estimators can be quantified and compared. For example, the relatively
large systematic estimation error in the implied timescales / implied
rates of Markov state models is explained. Additionally, the theory
provides a measure for the observation quality (OQ) $\hat{\alpha}_{o}$
of the observed signal, which is independent of any specific dynamical
model and also does not need the definition of an ``A'' or ``B''
state and bounds the error in rates estimated from the observed signal.
$\hat{\alpha}_{o}$ includes effects of the order parameter measured
as well as the effect of the experimental construct on the signal
quality, such as experimental noise. It is shown that $\hat{\alpha}_{o}$
is a lower bound to the true reaction coordinate quality (RCQ) due to choosing the order parameter, and
can thus be used as an indicator to both improve the quality of the
experimental setup and the choice of the order parameter.

The theory suggests steps to be taken to construct an optimal rate
estimator, that minimizes the systematic error in the estimation of
rates from an observed dynamical system. We propose such an estimator
and refer to it as \emph{spectral estimator}. It provides rather direct
and optimal estimates for the following three types of quantities:
\begin{enumerate}
\item The observation quality (OQ) $\hat{\alpha}_{o}$ of the observed
signal.
\item The dominant relaxation rate $\kappa_{2}$, as well as the microscopic
transition rates $k_{AB}$ and $k_{BA}$, even if $A$ and $B$ strongly
overlap in the observable.
\item The partial probability densities, and hence projections of the states
$A$ and $B$ in the observable, $\mu_{A}^{o}(o)$ and $\mu_{B}^{o}(o)$,
as well as their total probabilities, $\pi_{A}$ and $\pi_{B}$. This
information is also obtained if $A$ and $B$ strongly overlap in
the observable.
\end{enumerate}
Other rate estimators that rely on fitting the exponential tail of
a time-correlation function calculated from the experimental recorded
trajectories can also estimate $\kappa_{2}$ without systematic error.
However, the spectral estimator is unique in also being able to estimate
$k_{AB}$, $k_{BA}$, $\mu_{A}^{o}(o)$, $\mu_{B}^{o}(o)$ and the
OQ in the presence of states that overlap in the observable order
parameter.

\subsection*{Discussion}

The present study has concentrated on systematic rate estimation errors
that are expected in the data-rich regime. We expect that taking the
statistical error into consideration will make the spectral estimator
described here even more preferable over more direct approaches such
as fitting the number autocorrelation function of a dividing surface.
This intuition comes from the fact that the spectral estimator maximizes
the amplitude $\alpha$ with which the slow relaxation of interest
is involved in the autocorrelation function. In the presence of statistical
uncertainty, this will effectively maximize the signal-to-noise ratio
in the autocorrelation function and thus lead to an advantage over
fitting autocorrelation functions that were obtained differently.

Consideration of the statistical error will also aid in selecting an appropriate 
$\tau$ that balances systematic and statistical error in rate estimates.
$\tau$-dependent fluctuations of the sort observed in Fig.~\ref{fig:Rate-Estimate}III-5 might also be suppressed by averaging over multiple choices of $\tau$ in a manner that incorporates the statistical error estimates in weighting.

The presented idea of building an optimal estimator for a single relaxation
rate upon the transition matrix estimate of the projected slowest
eigenfunction, $\hat{\psi}_{2}$, is extensible to multiple relaxation rates, and this will be pursued in future studies. 
%This lays the basis for a multi-state rate theory
%for low-dimensional observations of dynamical systems that will be
%pursued in future studies. Such a theory will help us to understand
%and enhance the behavior of existing data analysis methods such as
%Hidden Markov models when applied to observed physical systems.

\subsection*{Acknowledgements}

The authors thank Susan Marqusee and Phillip J. Elms (UC Berkeley)
for sharing the single-molecule force probe data. FN and JHP acknowledge
funding from the DFG center \textsc{Matheon}. FN acknowledges funding from ERC
starting grant ``pcCell''. JDC acknowledges funding
from a QB3-Berkeley Distinguished Postdoctoral Fellowship during part of this work. We are
grateful to Christof Sch\"utte, Attila Szabo, Sergio Bacallado, Vijay
Pande and Heidrun Prantel for enlightening discussions and support.

\end{document}